%% file: main.tex
\documentclass[conference,compsoc]{IEEEtran}
\ifCLASSOPTIONcompsoc
  \usepackage[nocompress]{cite}
\else
  \usepackage{cite}
\fi
\ifCLASSINFOpdf
\else
\fi

\usepackage{hyperref}
\usepackage{multirow}
\usepackage{booktabs} 
\usepackage{epstopdf}
\usepackage{adjustbox}
\usepackage{tcolorbox} 
\usepackage{hhline}
\usepackage{pifont}

\usepackage{enumitem}

\usepackage{amsmath,amssymb,amsfonts}

\usepackage{graphicx}
\usepackage{textcomp}
\usepackage{xcolor}
\usepackage{tabulary}
\usepackage{colortbl}
\usepackage{float}
\usepackage{diagbox}
\usepackage{pifont}
\usepackage{bbding}
\def\BibTeX{{\rm B\kern-.05em{\sc i\kern-.025em b}\kern-.08em
    T\kern-.1667em\lower.7ex\hbox{E}\kern-.125emX}}

\usepackage{caption}
\usepackage{subcaption}
\usepackage{listings}
\usepackage{tikz}
\usetikzlibrary{arrows.meta}
\usepackage{color}
\usepackage{colortbl}

\usepackage{cuted}
\usepackage{xspace}
\usepackage{algorithm}
\usepackage{algpseudocode}

\usepackage{comment}
\usepackage{arydshln}
\usepackage{url}

\usepackage{dirtytalk}
\usepackage{tikz}
\usepackage{amsmath}

\usepackage{graphicx}

\newcommand{\rcpone}{\textit{Type-\uppercase\expandafter{\romannumeral1}}\xspace}
\newcommand{\rcptwo}{\textit{Buffered CRP}\xspace}
\newcommand{\rcpthree}{\textit{Scheduled CRP}\xspace}
\newcommand{\eg}{{\it e.g.,}}
\newcommand{\ie}{{\it i.e.,}}

\newcommand{\zw}[1]{\textcolor{blue}{#1}}
\newcommand{\sys}{{\textit{FIDO}}\xspace}
\newcommand{\CRProute}{{input route}\xspace}
\newcommand{\CRProutes}{{input routes}\xspace}
\newcommand{\uemu}{{\textit{$\mu$Emu}}\xspace}
\newcommand{\uafl}{{\textit{$\mu$AFL}}\xspace}
\newcommand{\semu}{{\textit{SEmu}}\xspace}

\newcommand{\fuzzware}{{\textit{Fuzzware}}\xspace}
\newcommand{\hoedur}{{\textit{Hoedur}}\xspace}

\newcommand{\mulitfuzz}{\textit{\normalsize M\small ULTI\normalsize F\small UZZ}\xspace}

\newcommand{\emberio}{{\textit{Ember-IO}}\xspace}
\newcommand{\aim}{{\textit{AIM}}\xspace}
\newcommand{\aid}{{\textit{AidFuzzer}}\xspace}
\newcommand{\ppim}{{\textit{P$^{2}$IM}}\xspace}

\newcommand{\CRP}{CRP\xspace}

\newcommand{\MSP}{{{MSP}}\xspace}

\definecolor{mGreen}{rgb}{0,0.6,0}
\definecolor{mGray}{rgb}{0.5,0.5,0.5}
\definecolor{mPurple}{rgb}{0.58,0,0.82}
\definecolor{backgroundColour}{rgb}{0.95,0.95,0.92}

\lstdefinestyle{CStyle}{
    commentstyle=\color{mGreen},
    keywordstyle=\color{magenta},
    stringstyle=\color{mPurple},
    basicstyle=\scriptsize\ttfamily,
    breaklines=true,
    captionpos=b,
    keepspaces=true,
    otherkeywords={status_t, ArrayObject,uint32_t},
    numbers=left,
    numbersep=1pt,
    showspaces=false,
    showstringspaces=false,
    showtabs=false,
    tabsize=2,
    language=C
}
\colorlet{punct}{red!60!black}
\definecolor{background}{HTML}{EEEEEE}
\definecolor{delim}{RGB}{20,105,176}
\colorlet{numb}{magenta!60!black}
\lstdefinelanguage{json}{
    basicstyle=\scriptsize\ttfamily,
    numbers=left,
    numberstyle=\scriptsize,
    stepnumber=1,
    numbersep=8pt,
    showstringspaces=false,
    breaklines=true,
    frame=lines,
    backgroundcolor=\color{background},
    literate=
      {:}{{{\color{punct}{:}}}}{1}
      {,}{{{\color{punct}{,}}}}{1}
      {\{}{{{\color{delim}{\{}}}}{1}
      {\}}{{{\color{delim}{\}}}}}{1}
      {[}{{{\color{delim}{[}}}}{1}
      {]}{{{\color{delim}{]}}}}{1},
}

\definecolor{codegreen}{rgb}{0,0.6,0}
\definecolor{codegray}{rgb}{0.5,0.5,0.5}
\definecolor{codepurple}{rgb}{0.58,0,0.82}
\definecolor{backcolour}{rgb}{0.95,0.95,0.88}

\definecolor{commentgreen}{HTML}{A3BE8C}    
\definecolor{bggray}{rgb}{0.95, 0.95, 0.95}
\definecolor{kwlavender}{rgb}{0.6, 0.4, 0.8}  
\definecolor{funcgreen}{rgb}{0.0, 0.5, 0.3}   
\definecolor{typecyan}{rgb}{0.0, 0.6, 0.8}    
\definecolor{linenumbergray}{rgb}{0.4, 0.4, 0.4}

\lstdefinelanguage{customC}{
  language=C++,
  morekeywords=[1]{if,else,for,while,return},
  morekeywords={[2]Modbus, UARTClass},
  morekeywords=[3]{uint32_t,int, uint8_t,char},
  sensitive=true,
  morecomment=[l]{//},
}

\newcommand{\cmtmathbox}[1]{{\color{commentgreen}\mbox{\(#1\)}}}

\AtBeginDocument{
  \setlength\abovedisplayskip{0pt}
  \setlength\belowdisplayskip{0pt}
  \setlength{\belowcaptionskip}{0pt}
 }
  \setenumerate[1]{itemsep=0pt,partopsep=0pt,parsep=\parskip,topsep=0pt}
  \setitemize[1]{itemsep=0pt,partopsep=0pt,parsep=\parskip,topsep=0pt}
  \setdescription{itemsep=0pt,partopsep=0pt,parsep=\parskip,topsep=0pt}

\renewcommand{\paragraph}[1]{\vspace*{0.04in}\noindent\textbf{#1}}

\begin{document}

\title{Stop Starving or Stuffing Me: Boosting Firmware Fuzzing Efficiency with On-demand Input Delivery}

\IEEEoverridecommandlockouts   

\author{
\IEEEauthorblockN{
Shandian Shen$^{*}$\thanks{
The first two authors contributed equally (alphabetical order).
},
Wei Zhou$^{*\text{\footnotesize\ding{41}}}$\thanks{Wei Zhou is the corresponding author.},
Keming Zhao$^{*}$,
Peng Liu$^{\dagger}$,
Chung Hwan Kim$^{\ddagger}$,
Le Guan$^{\S}$
}

\IEEEauthorblockA{
$^{*}$School of Cyber Science and Engineering, Huazhong University of Science and Technology, China \\
$^{*}$Hubei Key Laboratory of Distributed System Security \\ 
$^{\dagger}$Penn State University,
$^{\ddagger}$University of Texas at Dallas,
$^{\S}$University of Georgia
}

\IEEEauthorblockA{
E-mails:
\{shenshandian, weizhou\_sec, zhaokeming\}@hust.edu.cn,
pliu@psu.edu,
chungkim@utdallas.edu,
leguan@uga.edu
}
}

\maketitle

\begin{abstract}
Firmware fuzzing has gained attention for its ability to identify firmware bugs. 
While progress has been made in firmware emulation to support fuzzing, 
current approaches often directly integrate fuzzing tools for general software.
However, unlike general software, which receives input as it encounters I/O functions,
firmware input can be received asynchronously and independently of the firmware's execution, 
with uncertain \textit{timing and quantity}. 
Without full awareness of firmware's exceptions,
existing solutions often imprudently deliver fuzzer-generated input to the firmware in an ad-hoc way.
This either overwhelms the processing function of the firmware (i.e.,~\textit{stuffing} problem) or 
fails to deliver enough input data to trigger input processing functions (i.e.,~\textit{starving} problem).
In both cases, fuzzing capability is weakened.


In this paper, we comprehensively investigate the input delivery issue, a unique
and less studied field in firmware fuzzing.
To accurately determine the optimal timing and quantity for delivering test cases,
we leverage the fact that firmware has to check input availability before using any data.
Therefore, we employ static and dynamic analysis to map each input processing route
into three stages: input retrieval, availability check, and processing.
This recovered semantic information allows the fuzzer to accurately deliver input at the 
availability check points within the expected length range. 
Since firmware may have multiple input routes, we also optimize the scheduling algorithm
to reach more diverse routes. 
Our prototype, named \sys, can serve as an add-on to existing firmware fuzzers
to enhance their test-case delivery effectiveness.
Compared to ad-hoc input delivery methods used in \fuzzware and \mulitfuzz,
\sys increases their median code coverage by up to 115\% and 54\%, respectively. 
Compared to \semu, which requires humans to manually specify input delivery points,
\sys still improves its coverage by up to 19\%. 
As a result of improved input delivery strategy, 
\sys discovers known bugs significantly faster and
also identifies five previously unknown bugs.
\end{abstract}

\input{tex/1-intro}

\input{tex/2-background}

\input{tex/3-problem}

\input{tex/4.1-CRPmodel}

\input{tex/4.2-design}

\input{tex/5-impl}

\input{tex/6-eval}

\input{tex/7-discuss}

\input{tex/8-related_work}

\input{tex/9-con}

\section*{Ethics Considerations}

Our tool identifies vulnerabilities in MCU-based device firmware using fuzzing. We conduct experiments on firmware in emulators within an isolated internal server. 
All security bugs found in this work
have been reported to vendors/developers as detail in Appendix~\ref{app:bugstatus}.
The firmware images used in our study were sourced from public resources.

\section*{Acknowledgment}

We sincerely appreciate our shepherd and all the anonymous reviewers for their insightful and valuable feedback. This work was supported by National Natural Science Foundation of China (NSFC) grant (62202188).

\bibliographystyle{IEEEtranS}
\bibliography{bibs/main}

\appendices
\input{tex/11-app}

\input{tex/Meta-Review}

\end{document}

%% file: tex/1-intro.tex
\section{Introduction}

Microcontroller Unit (MCU) based embedded devices are widely used in security- and safety-critical sectors such as infrastructure, smart homes, and healthcare. 
At the same time, vulnerabilities in MCU-based devices have steadily risen.
Over the past decade, fuzzing has emerged as a highly effective technique for detecting software vulnerabilities.
Therefore, researchers have been working on adopting traditional fuzzers 
for firmware testing which relies on real  hardware~\cite{li2022muafl,mera2024shift,zaddach2014avatar,muench2018avatar2} 
or an emulator~\cite{zhou2021automatic,zhou2022your,tobias2022fuzzware, spensky2021conware,won2022what,chong2024afriend} for code execution.
Beyond traditional fuzzing, recent studies also
retrofit fuzzing strategies to accommodate firmware-specific features~\cite{scharnowski2023hoedur,chesser2024multifuzz,farrelly2023splits}.
For example, the authors of \hoedur~\cite{scharnowski2023hoedur} and \mulitfuzz~\cite{chesser2024multifuzz} found that firmware receives inputs from multiple peripheral interfaces.
They correspondingly propose to maintain a separate input stream
for each peripheral and independently mutate them.
Such multi-stream fuzzing reorganizes the previously series single-stream fuzzing input into per-peripheral streams that are independently mutated,
enabling peripheral-specific and type-aware fuzzing.


\paragraph{Input Delivery Problem in Firmware Fuzzing.}
In this paper, we demonstrate that being orthogonal to the multi-stream problem, there is another equally critical problem (i.e., the input delivery problem) caused by firmware specificities: solving this problem can significantly boost fuzzing performance.

We now explain the input delivery problem in firmware fuzzing if this specificity is not handled properly.
Traditional software uses well-defined POSIX I/O interfaces
such as \texttt{read(fd, buf, len)} to synchronously receive input from files or keyboards.
Conventional fuzzers simply hook into these I/O APIs to deliver inputs to the corresponding \texttt{fd}, ensuring inputs are always available without needing to manage delivery quantity, as the program's function parameters dictate the expected input byte amount. 
In contrast, firmware retrieves and processes input from various peripherals with \emph{non-standard} I/O interfaces and
\emph{asynchronous} triggering mechanisms (\eg~interrupt), 
making the \emph{arrival time and quantity} of peripheral input largely unpredictable.

\begin{figure}[t]
\setlength\abovedisplayskip{0pt}
\setlength\belowdisplayskip{0pt}
\centering
\includegraphics[width=\columnwidth]{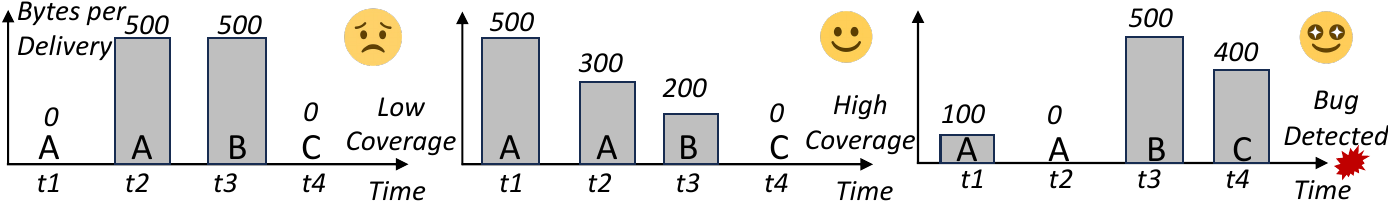}
\caption{Three delivery plans with varying timing and quantities (\texttt{A}, \texttt{B}, and \texttt{C} refer
to different input-taken points).}
\label{fig:intro}
\end{figure}

Without the knowledge of expected input arrival time and quantities, existing fuzzers make ad hoc input delivery decisions/plans and fail to optimally deliver input,
often leading to low fuzzing effectiveness and efficiency.
We illustrate three possible input delivery plans in~\autoref{fig:intro}.
In each plan, assuming that in total 1,000 bytes are delivered to three input-taken points \texttt{A}, \texttt{B}, and \texttt{C}.
A fuzzer decides whether to deliver input and how many bytes at each time slot \textit{t1} to \textit{t4} when encountering these points.
0 bytes means that the fuzzer determines not to deliver any input at that time slot. 
Depending on the firmware, these plans have led to very different code coverage and bug detection outcomes.
The first plan fails; the second plan increases code coverage; the third plan detects a bug.


\paragraph{Limitations of Existing Works.}
Existing firmware fuzzers deal with input delivery in ad-hoc ways, leading to suboptimal efficiency. 
For instance, \fuzzware, \hoedur, and \mulitfuzz support two fuzzing modes: periodic and fuzz.
In the periodic model,
a few bytes are delivered by triggering interrupts at fixed intervals (\eg~every 1,000 basic blocks).
The fuzz mode is similar to the periodic mode, except that the interval
and interrupt are dynamically determined at runtime based on the fuzzing input.
These earlier works, without understanding firmware's expectation,
are more likely to inject extra bytes when the firmware does not expected any,
or fail to deliver bytes when the firmware really needs data.
\semu relies on manually specified delivery points, but does not account for the size limit of peripheral buffers.
This causes the old buffer to be overwritten, thus wasting test cases.
Additionally, manual configuration is also labor-intensive and may not capture all the valid input delivery points, as we demonstrate in evaluation.

A concurrent work, \aid~\cite{wangaidfuzzer}, aims to determine when firmware is awaiting interrupts by identifying waiting states like infinite loops or points reading global variables influenced by interrupt handlers. The heuristic is that when the firmware communicates with the interrupt handler, which handles low-level data reception, it expects new data. If such a state is encountered a specified number of times, \aid triggers an interrupt.
However, these waiting states do not necessarily represent the firmware's genuine intention to consume input.
As a result, this coarse-grained heuristic often over-approximates the inference problem, 
flagging cases where no input is actually required.
We discuss problems with the coarse-grained heuristic in evaluation (\autoref{sec:eval:aid}).
\paragraph{Our Solution.} 
Our work makes three key contributions to facilitate more efficient and effective firmware fuzzing.
First, we propose a new approach to more precisely identify input retrieval points.
Rather than inferring waiting states through accesses to shared variables, 
our approach identifies the complete control-flow route along which input is Checked (for availability),
Retrieved, and subsequently Processed---a sequence  (of operations) we term a ``\textbf{CRP input route}''.
For each such route, input is delivered directly at the check locations.
This semantic-level reasoning provides a more reliable indication that the firmware genuinely requires input.
Second, our approach can determine the range of input sizes that the firmware can effectively process.
The proposed ``\textit{watermark}'' technique gradually increases input length
until the firmware begins to process input, indicating the lower limit.
The upper limit is detected when the delivered input continues growing and leads to
the old data being overwritten in the receive buffer.
Lastly, we propose a multi-route-aware input delivery strategy.
Firmware can have multiple \CRProutes, each with varying input length requirements and 
occurring under different calling contexts. 
These \CRProutes may be hit at different times during fuzzing. 
Without coordinating input delivery for these routes, 
most inputs would be directed to a small group of routes, leaving others less tested.
The proposed strategy dynamically predicts potential input routes and distributes test cases fairly among all \CRProutes.

We have implemented our idea with a prototype named \sys (acronym for \underline{F}uzzing \underline{I}nput \underline{D}elivery \underline{O}ptimizer).
We have integrated \sys as an input delivery plugin with three state-of-the-art (SOTA) firmware fuzzers:
\fuzzware~\cite{tobias2022fuzzware}, \semu~\cite{zhou2022your}
and \mulitfuzz~\cite{chesser2024multifuzz}, replacing their original input delivery mechanisms.
Based on our evaluation of 28 unit tests and 25 real-world firmware images, 
\sys accurately identified optimal delivery strategies for all samples.
With more efficient input delivery, \sys reduces input wastage and increases likelihood of passing availability checks.
This improvement translates to significantly boosted fuzzing capability.
In our five groups of 24-hour fuzzing campaigns for real-world firmware images,
\sys increases median code coverage by up to 115\% and 54\% compared to the periodic 
delivery methods of \fuzzware and \mulitfuzz, respectively,
and by up to 106\% compared to \fuzzware's fuzzed mode. 
Compared to SEmu, which requires humans to manually specify
input delivery points, \sys still improves its coverage by up
to 19\%.
\sys enhances fuzzers' bug-finding capabilities, 
triggering crashes 1 to 100 times more often than these with the original ones, 
and discovering known bugs hundreds of times faster.
With the help of \sys, we found five new bugs and one of them was assigned with CVE number.





In summary, our contributions are four-fold:
\begin{itemize}
\item We reveal the test case delivery problem, a firmware-specific roadblock hindering efficient firmware fuzzing. 

\item We address the test case delivery problem using \CRProute analysis. 
By abstracting an \CRProute with the proposed \CRP model (availability \textbf{C}heck, data \textbf{R}etrieval, \textbf{P}rocessing), 
we can understand when the firmware really needs input and how many bytes to deliver.

\item We implement our idea as a plugin called \sys for four SOTA firmware fuzzers.

\item Our evaluation results confirm that \sys benefits all the four SOTA firmware fuzzers, 
effectively improving test case usage, increasing code coverage, and triggering more crashes within shorter time.

\end{itemize}

To facilitate further research, we have released the \sys source code and dataset at \url{https://github.com/IoTS-P/FIDO}.

%% file: tex/2-background.tex
\section{Background}

\label{sec:background}
\subsection{MCU Firmware and Peripheral Interface}
Firmware is a program dedicated for an embedded device.
When running on an MCU-based embedded device,
it is often monolithic, comprising application code, drivers, libraries, and an optional real-time kernel altogether.
If a real-time kernel is included,
the firmware is said to be RTOS-based; otherwise, it is said to be bare-metal.
Since MCU firmware is dedicated for a particular task,
it works in an infinite loop that
continuously receives input data from the external world, processes it, and responds to the external world via actuators when necessary.

MCU firmware lacks a unified machine abstraction layer, requiring direct interaction with peripherals through low-level hardware interfaces. There are three main I/O mechanisms in MCUs. First, Memory-Mapped I/O (MMIO) maps peripheral registers into the system address space, allowing firmware to read or write data registers (DR) for input or output. Second, peripherals can signal status changes via Interrupt Requests (IRQs), managed by an interrupt controller like the nested vector interrupt control (NVIC) in ARM Cortex-M MCUs. Third, peripherals such as USB and Ethernet use Direct Memory Access (DMA) for bulky data transfers between RAM and peripherals without processor intervention, enhancing throughput.

\subsection{I/O Operating Modes on MCUs}

\begin{figure}[t]
\setlength\abovedisplayskip{0pt}
\setlength\belowdisplayskip{0pt}
\centering
\includegraphics[width=\columnwidth]{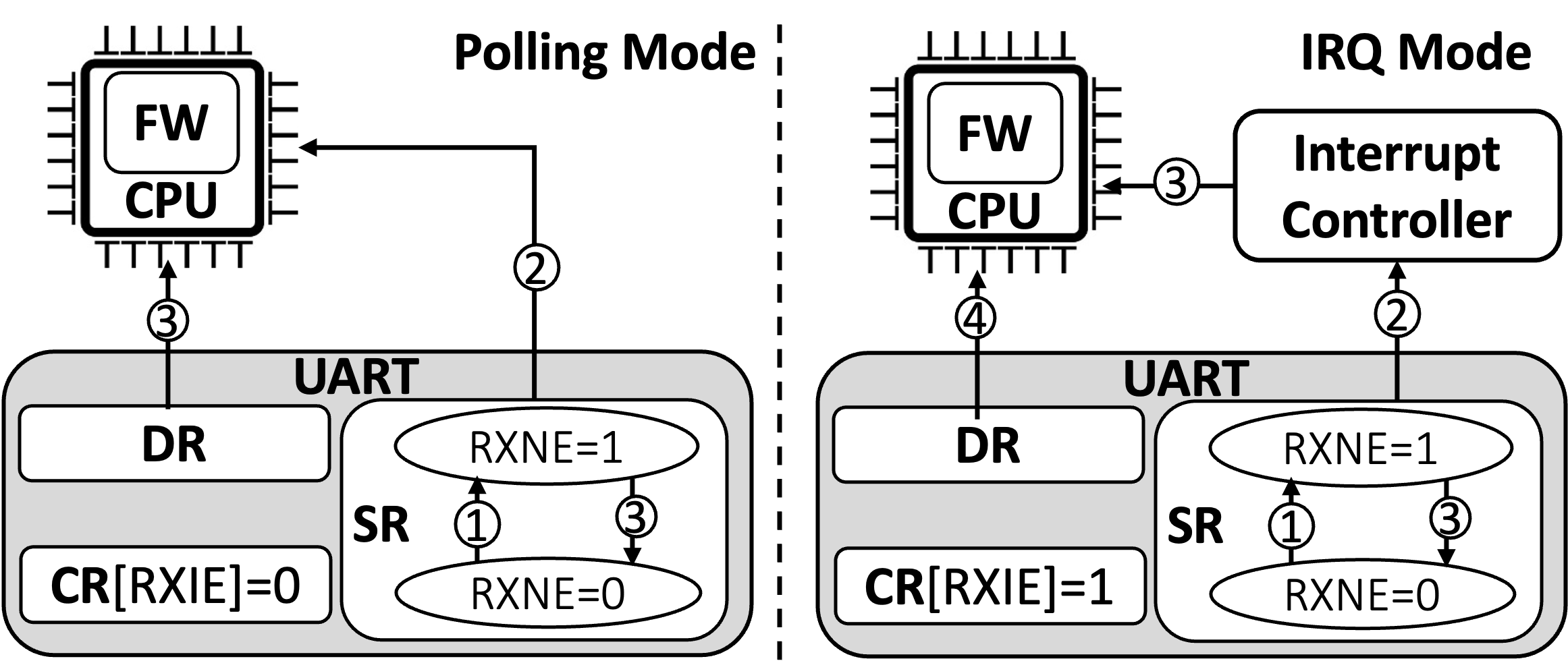}
\caption{Two I/O operating modes for peripheral input arrival. FW refers to firmware.}
\label{fig:iomodes}
\end{figure}

The firmware retrieves external input from peripherals either via reading data register or DMA transfers. 
Since peripheral hardware operates asynchronously with the processor core, 
it must notify the firmware of input arrival before reading it. 
The notification is achieved either passively by letting the 
firmware poll the status register (polling mode), 
or actively by triggering an interrupt to the current execution (IRQ mode).
Using two simple unit tests of UART peripheral on STM32F429 as an example, 
we demonstrate the workflows of these two operating modes in~\autoref{fig:iomodes}.



\paragraph{Polling Mode.} When external data reaches a peripheral, it updates the status registers (SR). 
In the UART example, this corresponds to setting the \texttt{RXNE} field of the SR, 
indicating that the peripheral is ready to be read (\ding{172}).
Before accessing the data register (DR), the firmware must check the \texttt{RXNE} field (\ding{173}). 
The read operation can continue until the \texttt{RXNE} field is 
cleared (\texttt{Empty} state) by the peripheral, which means that the input is read out (\ding{174}).



\paragraph{IRQ Mode.}
Polling mode inefficiently uses CPU cycles for input availability checks. 
This can be mitigated by the interrupt mechanism which allows the hardware to automatically notify firmware of data arrival.
In the UART example, the firmware can enable interrupt by setting corresponding bit in 
the \texttt{RXIE} field in the control register (CR). 
When UART input arrives, not only the \texttt{RXNE} field in SR is set (\ding{172}), 
but also the corresponding peripheral interrupt is activated by the interrupt controller (\ding{173}). 
This automatically switches the processor from its current execution 
context (thread mode in ARM Cortex-M) to a predefined interrupt service routine (ISR) 
context (handler mode in ARM Cortex-M) (\ding{174}) to read the data.
When the peripheral input is read out, the
peripheral turns to the \texttt{Empty} state (\ding{175}).
The interrupt mechanism eliminates the need for constant peripheral checks
at the cost of increased software complexity to deal with asynchrony (detailed later in~\autoref{sec:programmodel}).

\subsection{Firmware Specificity involved in the Programming Model}
\label{sec:programmodel}

Special programming models are employed to accommodate the aforementioned I/O operating modes on MCUs,
as recognized in the community~\cite{wikiiomode}.
In the polling mode, the firmware constantly checks readiness via reading the SR.
Once the condition is met, the firmware retrieves and processes the input in the same execution context.
When operating in the IRQ mode, 
the firmware is free to handle other tasks until it is notified of data arrival asynchronously by the interrupt controller.
Data is then typically retrieved in the ISR context promptly
and the processing is put off to a later stage outside of the ISR context.
This work focuses on fuzzing firmware using the asynchronous IRQ mode
because existing firmware fuzzers can already handle firmware in
the polling mode that synchronously receives and processes input.
In what follows, we summarize four unique characteristics caused by the asynchronous IRQ mode.
They directly impact the design of the proposed input delivery strategy.



\paragraph{F1: Limited Reserved Buffer for Input Retrieval.}
In the IRQ mode, an ISR function is activated upon input arrival, which runs asynchronously with the main firmware function.
ISRs are short-lived routines and must exit as soon as possible~\cite{israsap}.
Therefore, a global buffer is commonly reserved to allow the ISR to quickly receive
the data from peripheral and exit. 
Later, the main function outside the ISR context can process it as needed.
Each peripheral in IRQ mode requires at least one dedicated buffer.
The size of the shared buffer varies depending peripheral usage, 
generally ranging from several bytes to under 1 KB, due to limited SRAM resources in MCUs.
Furthermore, to ensure that the mostly recent bytes are still available in case of data burst,
the buffer is typically arranged as a ring data structure.
That is, when the buffer reaches its limit, the ISR overwrites the oldest bytes with the newly received data.

\paragraph{F2: Buffer Cleanup at Peripheral Reset.}
Every time when the peripheral needs to be reset into a predictable and clean state,
the associated ISR buffer is cleared.
This can occur during device reboot
or when the peripheral switches between the receiving and transmitting modes in case of shared receiving and transmitting buffer.
Obviously, unprocessed data in the buffer during cleanup will be discarded and cannot contribute to the fuzzing process.

\paragraph{F3: Availability and Length Check.}
Since the ring buffer is filled by ISR functions asynchronously, 
the firmware must check if there is new data in the buffer before using it.
Additionally, some protocols require a minimal amount of data before they can be processed.
The firmware thus also needs to perform a length check of data.
Data availability check is implemented in a more principled way when the firmware incorporates an RTOS kernel.
Specifically, each input source is assigned with a semaphore and each task monitors a set of semaphores.
If no semaphore is available for a task,
the task is moved to a sleep list. 
If a semaphore is released from ISRs,
the task waiting on this semaphore wakes up.
If no task is ready to run, the system defaults to an idle task.


\paragraph{F4: Multiple Processing Points with Different Calling Contexts.}
During firmware execution, the input processing functions may be invoked at multiple places in different contexts.
The number of consuming points encountered can vary depending on the input data. 
Additionally, the behavior of the processing function may change according to the calling context.



\paragraph{Generality Study.}
To assess the generality of these characteristics in real MCU firmware,
we examined 87 unit test demos shipped with official SDKs of top MCU vendors as listed in Appendix~\ref{app:feature}, 
including STM32, NXP, and Microchip, and popular MCU driver library Arduino
as well as leading RTOSs like FreeRTOS, Zephyr, Mbed OS, RIoT, and Nuttx, covering
commonly-used peripherals such as GPIO, I2C, UART, ADC, SPI.
Among the 87 unit test samples, 54 incorporate F1. 
The remaining samples that do not implement F1 primarily utilize 
very simple peripherals for direct data transfer, thereby eliminating the need for a ring buffer in RAM. 
Exceptions include certain samples involving GPIO, and ADC.
Due to the inherently asynchronous interaction between peripherals and firmware, 
all samples incorporate F3. 
No unit test was found to implement F4, 
as these tests are specifically designed to validate minimal, isolated functionalities of individual peripherals.
In contrast, our analysis of the real-world firmware dataset shows that 19 out of 25 samples include F4.

\subsection{CRP Modeling for MCU Firmware}

According to the programming model, 
we found that the operational process firmware uses to handle an input consists of three steps: \textit{availability check, retrieval, and processing (CRP).}
For clarity, we define the CRP operations in this work as follows:

\begin{itemize}

\item \textbf{Availability Check ($C_{A}$).}
In the polling mode, the firmware directly checks if the SR field which indicates 
readiness (e.g., the \texttt{RXNE} field in UART SR is set in~\autoref{sec:background}).
In the IRQ mode, 
the main firmware logic checks the buffer populated by the ISRs.
In RTOS-based firmware, the semaphore of the input source is checked.
In addition, some firmware also performs a minimum length check on the accumulated data in the buffer, denoted as $C_{L}$.

\item \textbf{Data Retrieval ($R$).}  In the polling mode, after confirming the SR field,
the firmware reads the DR as data retrieval. 
In the IRQ mode, data retrieval occurs at two levels.
First, the ISR reads from the peripheral's data I/O and stores it in the buffer (low-level data retrieval, $R_{P}$).
Second, after the low-level availability check, 
the main firmware logic retrieves data from the ISR buffer (high-level data retrieval, $R_{B}$).
If $R_{B}$ occurs during buffer cleanup (F2), it is denoted as $\widetilde{R_{B}}$.

\item \textbf{Processing ($P$).} In both polling and IRQ modes, the retrieved
data is processed by the main logic of the firmware.
Operations involving the input, particularly those influence the branch targets at conditional jumps, 
can contribute substantially to code coverage.
These operations drive the main application logic and form the core of the firmware code.

\end{itemize}

In IRQ mode, an input handling process creates a distinct sub-tree in the control flow graph, connecting instructions of $C_{A}$, $R_{B}$, and processing ($P$), referred to as an \CRProute in this paper.

%% file: tex/3-problem.tex
\section{Motivation}

\subsection{Input Delivery Problem}
\label{sec:fuzzdelivery}

As with general software,
the firmware fuzzing loop involves five modules:
test case generation, test case delivery, execution
with the test case, feedback collection, and feedback-guided input mutation.
In this work, we refer to the firmware executor as the fuzzer back-end, 
and the test case generation and mutation module as the fuzzer front-end.
Due to architectural differences,
rehosting has been widely adopted as the fuzzer back-end (\eg~\uemu~\cite{zhou2021automatic}, \semu~\cite{zhou2022your} and \fuzzware~\cite{tobias2022fuzzware})
in firmware fuzzing, where the firmware instructions are translated into host instructions.
Regarding the fuzzer front-end, previous works typically adopt test case
generation algorithms shipped with AFL/AFL++ or libfuzzer. 


Recently, researchers~\cite{scharnowski2023hoedur, chesser2024multifuzz, farrelly2023splits} found that firmware exposes many 
specificities that are incompatible with existing fuzzing front-ends. 
For example, multi-stream works such as \mulitfuzz and \hoedur enhance input generation and mutation to handle the multi-stream nature of firmware inputs from various peripherals.
However, when it comes to input delivery (i.e., when and how many bytes to deliver), 
we found that existing firmware fuzzers overlook the unique programming model in handling inputs in the IRQ mode, and instead use ad-hoc mechanisms, as summarized in~\autoref{tab:fuzzdelivery}.
For instance, most emulation-based firmware fuzzers use a round-robin injection method (\textbf{RR}), 
delivering input by triggering an ISR after every fixed number (\eg~1,000) of basic blocks executed.
The input quantity per trigger is usually between 1B and 4B, as requested in each ISR.
Additionally, \fuzzware, \hoedur and \mulitfuzz supports a fuzzing-guided delivery method (\textbf{Fuzz}), 
where input is delivered with a single ISR trigger but at different intervals determined by the fuzzing input.
If multiple peripheral interrupts are enabled, the interrupt chosen is cycled in RR mode and is determined by fuzzing input in Fuzz mode.
In comparison,
high-fidelity emulation-based fuzzers like \semu and on-device fuzzers like \uafl 
deliver all fuzzing input at manually specified points (\textbf{\MSP}), typically at the start of the main loop.

\begin{table}[t]
\caption{Front-end, back-end and input delivery method supported by existing firmware fuzzers for IRQ mode. (with default configurations in bold)}
\centering
\begin{adjustbox}{max width=\columnwidth}
\begin{tabular}{l|l|l|c|c}
\hline
\multicolumn{1}{c}{\textbf{Firmware}}          & \textbf{Front-end} & \textbf{Back-end}               & \textbf{Delivery}      & \textbf{Delivery Quantity} \\
\multicolumn{1}{c}{\textbf{Fuzzer}}            & \textbf{(Fuzzer)}  & \textbf{(Executor)}             & \textbf{Timing}        & \textbf{Each Time}     \\ \hline \hline
\ppim~\cite{fengp2020p2im}                     & AFL                & QEMU                            & \textbf{RR}           & Single ISR Read        \\
\uemu~\cite{zhou2021automatic}                 & AFL                & S2E                             & \textbf{RR}, \MSP       & Single ISR Read        \\
\emberio~\cite{farrelly2023ember}              & AFL++              & QEMU                            & \textbf{RR}            & Single ISR Read        \\
\fuzzware~\cite{tobias2022fuzzware}            & AFL, AFL++         & Unicorn                         & \textbf{RR}, Fuzz, \MSP & Single ISR Read        \\
\semu~\cite{zhou2022your}                      & AFL, AFL++         & Unicorn                         & \textbf{\MSP}           & All Fuzzing Input              \\
\hoedur~\cite{scharnowski2023hoedur}           & LibFuzzer*         & QEMU                            & \textbf{RR}, Fuzz, \MSP       & Single ISR Read        \\
\textit{MultiFuzz}~\cite{chesser2024multifuzz} & AFL*               & Icicle~\cite{chesser2023icicle} & \textbf{RR},Fuzz            & Single ISR Read        \\
\uafl~\cite{li2022muafl}                       & AFL                & Hardware                        & \textbf{\MSP}           & All Fuzzing Input              \\ 
\hline
\end{tabular}
\end{adjustbox}
\label{tab:fuzzdelivery}
\caption*{%
    \scriptsize
    \raggedright
*: Support multi-stream test case generation and mutation.
}
\end{table}

\subsection{How Ad-hoc Solutions Fall Short}
\label{sec:problems}

In this section, we analyze how existing delivery methods \textit{stuff} (problems P1, P2) and \textit{starve} (problems P3 and P4) firmware execution, leading to inefficient fuzzing with real-world firmware (see \autoref{lst:heatpress}).
This firmware operates on a \texttt{Heat\_Press} device, which receives 
remote commands via the Modbus protocol over a UART peripheral in IRQ mode and 
exhibits all the four features mentioned in~\autoref{sec:programmodel}.

\begin{lstlisting}[label={lst:heatpress},
abovecaptionskip=0pt, belowcaptionskip=0pt, escapechar=|,
caption={A simplified code snippet of the Heat Press firmware is provided, with key operations commented in the corresponding lines.}]
//ISR Context
void UART_IRQHandler() {
    store_char(&Serial, Serial->UART_DR); 
}
void store_char(RingBuffer *this, char c) {
    if ((this->Head + 1) & 127 != this->Tail) {
        this->rx_buffer[this->Head] = c;|\label{line:retrival1}| //|\cmtmathbox{R_{P}}| 
        this->Head = (this->Head + 1) & 127;|\label{line:lenmax}|
    }
}
// Main Execution Context
void main(...) {
  ...
  Modbus::begin(&slave, ...);
  while (1) {loop();}
}
void Modbus::begin(Modbus *const this, ...) {
     ...
    while (this->Head - this->Tail != 0) |\label{line:voidcheck1}|
        this->port->read(); |\label{line:voidread1}|
}
int UARTClass::read(UARTClass *this) {   
     result = this->rx_buffer[this->Tail];|\label{line:retrival2}|//|\cmtmathbox{R_{B}}|
     this->Tail = (this->Tail + 1) & 127;
     return result;
}
void loop(...) { |\label{line:loop}|
    Modbus::query(&master, x);
    switch(slave.state) {
        case 1: Modbus.poll(); |\label{line:poll1}|
            if (master.state == 0) {
                ...|\label{line:process1}|//|\cmtmathbox{P}|
                slave.state++;|\label{line:slaveinc}|
                master.state = 1; |\label{line:master1}|
            }
        case 2: Modbus.poll(); |\label{line:poll2}|
            if (master.state == 0) {
              ...|\label{line:process2}|//|\cmtmathbox{P}|
}
void Modbus::query(Modbus *const this, ...) {
    while (this->Head - this->Tail != 0) |\label{line:voidcheck2}|
        this->port->read(); |\label{line:voidread2}|
    ... //data transmission
}
int Modbus::poll(Modbus *const this, ...) { |\label{line:poll}|
    if (this->port->available() == 0)|\label{line:check1}|//|\cmtmathbox{C_{A}}|
        return 0;
    if (Modbus::getRxBuffer(this) <= 7)|\label{line:lenmin}|//|\cmtmathbox{C_{L}}|
        return 1; 
    if (au8Buffer[1] & 0x80)|\label{line:process0}|//|\cmtmathbox{P}|
    ....
    master.state = 0; |\label{line:master0}|
}
int UARTClass::available(UARTClass *this) { 
    return this->Head - this->Tail & 127;  |\label{line:head}|
}
int Modbus::getRxBuffer(UARTClass *this) {
    BufferSize = 0;
    while (this->port->available()) {//|\label{line:check2}||\cmtmathbox{C_{A}}|
        auBuffer[BufferSize] = this->port->read();
        BufferSize++;
    }
    return BufferSize;
}
\end{lstlisting}


\paragraph{P1: Input Overwritten with Overfeeding.}
As F1 states, peripheral inputs are asynchronously retrieved into a global buffer by the ISR, and later processed by the main function.
If an excessive number of bytes are delivered to the buffer and the main function fails to process them promptly, 
the buffer may become full, leading to either the loss of incoming data or the overwriting of previously stored bytes.
In the example, the ISR function calls \texttt{store\_char} to read a byte from the DR and store it in \texttt{rx\_buffer}. If more data remains, the ISR is triggered again,
accumulating data in the \texttt{rx\_buffer}.  
If input exceeds 127 bytes, the \texttt{Head} pointer is reset to zero (\autoref{line:lenmax} in~\autoref{lst:heatpress}).

This overfeeding issue is common in fuzzers using the MSP delivery method, which delivers all input at once.
Firmware fuzzers, however, do not know the buffer's maximum limit and can generate extremely long data
in mutation, especially in the havoc process.
Thus, for this example, when fuzzer-generated test case length over 127 the out-length data will overwritten the before one.
This issue also occurs with RR and Fuzz delivery methods during rapid deliveries when $R_{B}$ is infrequent, but the ISR trigger interval is short. 
In both cases, the high-level consumer ($R_{B}$ and $P$) fails to poll all the received data by $R_{P}$ in time causing input wastage.
Overfeeding not only wastes fuzzing input but also causes the fuzzer to spend significant time mutating unused inputs.

\begin{table}[t]
\caption{Comparison testing under different delivery configurations. (The bold line indicates the default bytes per delivery used by \fuzzware.)}
\centering
\begin{adjustbox}{max width=\columnwidth}
\begin{tabular}{cc|ccc|cc|c|l}
\hline
\multicolumn{2}{c|}{\textbf{Delivery Method}}                                                    & \multicolumn{3}{c|}{\textbf{\# BB Coverage}}   & \multicolumn{2}{c|}{\textbf{Vol. Avg.}} & \multirow{2}{*}{\textbf{\begin{tabular}[c]{@{}c@{}}Time\\ Avg.(s)\end{tabular}}} & \multirow{2}{*}{\textbf{Problem}} \\
\textbf{Timing}                                                    & \textbf{Bytes per Delivery} & \textbf{Min.} & \textbf{Max.} & \textbf{Avg.}  & \textbf{Retriev.}    & \textbf{Proc.}   &                                                                                  &                                   \\ \hline
\multirow{3}{*}{RR}                                       & \textbf{1}                  & \textbf{486}  & \textbf{498}  & \textbf{491}   & \textbf{1000}        & \textbf{861}     & \textbf{15.66}                                                                   & \textbf{P2-4}                   \\
                                                                   & Rand(0-1000)              & 410           & 422           & 418            & 245                  & 86               & 0.71                                                                             & P1-4                          \\
                                                                   & Rand(7-128)               & 455           & 461           & 457            & 982                  & 724              & 0.73                                                                             & P2-4                            \\ \hline
\multirow{3}{*}{Fuzz}                                     & \textbf{1}                  & \textbf{472}  & \textbf{502}  & \textbf{485}   & \textbf{1000}        & \textbf{887}     & \textbf{5.67}                                                                    & \textbf{P2-4}                   \\
                                                                   & Rand(0-1000)              & 420           & 438           & 429            & 369                  & 102              & 0.61                                                                             & P1-4                          \\
                                                                   & Rand(7-128)               & 457           & 461           & 459            & 984                  & 652              & 0.65                                                                             & P2-4                            \\ \hline
\begin{tabular}[c]{@{}c@{}}\MSP\\ (loop)\end{tabular}              & Rand(0-1000)              & 457           & 504           & 483            & 1000                 & 1000             & 0.79                                                                             & P1,3                              \\ \hline
\textbf{\begin{tabular}[c]{@{}c@{}}Ideal\\ Placement\end{tabular}} & \textbf{Rand(7-128)}      & \textbf{493}  & \textbf{516}  & \textbf{503}   & \textbf{1000}        & \textbf{1000}    & \textbf{0.68}                                                                    & \textbf{-}                        \\ \hline
\end{tabular}
\end{adjustbox}
\label{tab:motivation}
\flushleft
\end{table}

\paragraph{P2: Input Discard Due to Incorrect Timing.}
As mentioned in F2, data retrieved via DR is discarded during buffer cleaning ($\widetilde{R_{B}}$). In the example, the \texttt{Modbus.begin} function clears the current UART input buffer by retrieving all data without using or storing it at~\autoref{line:voidread1}. Similarly, \texttt{Modbus.query} retrieves buffer data before transmission.

The firmware fuzzer cannot differentiate between data from $\widetilde{R_{B}}$ and normal $R_{B}$ as they use the same instruction. 
If RR or Fuzz mode provides input bytes before \texttt{Modbus.begin} or \texttt{Modbus.query}, these bytes are wasted, leading the real processed volume of input can be smaller than the volume retrieved. 
A similar issue can arise with \MSP when the specified delivery points before $\widetilde{R_{B}}$.
Identifying all $\widetilde{R_{B}}$ requires significant manual effort, as it can occur under specific conditions in the firmware logic, as seen in \semu configurations where delivery happens after the \texttt{Modbus.begin} function but the delivery point is incorrectly set before the \texttt{query} function.





\paragraph{P3: Unnecessary Availability \& Length Check.}
If no available data to main input source, firmware would repeat availability checking, wasting execution time.
In the example, the \texttt{main} function runs an infinite \texttt{loop()} that calls the \texttt{Poll} function. 
This function uses \texttt{available} to check input availability by comparing the buffer's head and tail pointers. 
In addition to availability, \autoref{line:lenmin} also ensures that the data length meets the Modbus protocol's minimum 
requirement (\textgreater7), before processing the data in \texttt{auBuffer}.
The RR and Fuzz methods do not specify delivery timing, making it difficult to timely delivery before availability checks.
This limitation becomes particularly pronounced in scenarios where 
only few bytes is transmitted as single ISR trigger (which is default configuration) per delivery and additional length check is needed.

\paragraph{P4: Difficulty in Exploring New Behaviors.}
As described in F4, the firmware retrieves data from different program points at various times. 
If the fuzzer fails to deliver input at certain points, some availability checks may fail, missing processing functions. 
In this firmware, although the \texttt{Poll} function only retrieves data from UART, 
the \texttt{Poll} function is called in multiple switch-case branches under different contexts.
If all the input are delivered to a fewer cases, some cases will have no data to retrieve, causing certain processes to be skipped.
RR and Fuzz is hard to delivery the input in time for all these case in random pattern.
For the \MSP, 
delivering data only once at beginning of \texttt{loop} execution ensures that the \texttt{Poll} function 
has input at~\autoref{line:poll1},
but leaves no data for the remaining \texttt{Poll} invocations (\eg~\autoref{line:poll2}).
Consequently, the processing code starting at~\autoref{line:process2} is not executed. 

\paragraph{Key Idea of Optimal Delivery.}
To address the identified issues, the optimal delivery time and quantity should meet the following requirements:

\begin{itemize}
    \item Input delivery timing should occur only at the initial input availability checkpoint of each \CRProute. This prevents issues P2 (where data retrieved by $\widetilde{R_{B}}$ cannot be processed and included in an \CRProute), P3, and P4.
    \item The length of delivered input bytes should remain below the ISR buffer's maximum capacity to avoid P1.
    \item Delivery input bytes must satisfy the minimum length requirement and ensure that input exceeds this length with each delivery to prevent P3.
\end{itemize}


\paragraph{Delivery Strategy Comparison Demo.}
To verify our idea and assess how different input delivery strategies influence fuzzing progress, we fuzzed the firmware in~\autoref{lst:heatpress} using \fuzzware, which supports RR, Fuzz and \MSP delivery strategies. We extended it to include waiting state monitoring for the wait delivery strategy.
The demo runs each strategy using an identical batch of five 1,000B random inputs.
This simulates a typical firmware fuzzing process in which the firmware runs indefinitely
until all the input bytes are consumed with injected interrupts.
The demo also runs a manually specified ideal configuration according to our idea,
where inputs are delivered at the first \texttt{available} function invocation in the \texttt{Poll} function. Delivery quantity varied among the four strategies. The ideal strategy delivered 7 to 128 bytes each time to meet firmware requirements. Detailed configurations are in Appendix~\ref{app:config}.

We compared average time consumption, total retrieved data (via \texttt{rx\_buffer}), total processed data (bytes filled into \texttt{auBuffer}), and code coverage. As shown in~\autoref{tab:motivation}, the ideal strategy (with manual effort) achieved the highest code coverage, no input wastage, and nearly the shortest time.
The other tested strategies cannot 
1) guarantee that the retrieved and processed data volume matches the delivery volume,
2) limit execution time, 
or 3) cover all input retrieval/processing points simultaneously.
This work aims to achieve similar results as this ideal configuration via automated program analysis.

%% file: tex/4.1-CRPmodel.tex
\begin{figure}[t]
\setlength\abovedisplayskip{0pt}
\setlength\belowdisplayskip{0pt}
\centering
\includegraphics[width=0.9\columnwidth]{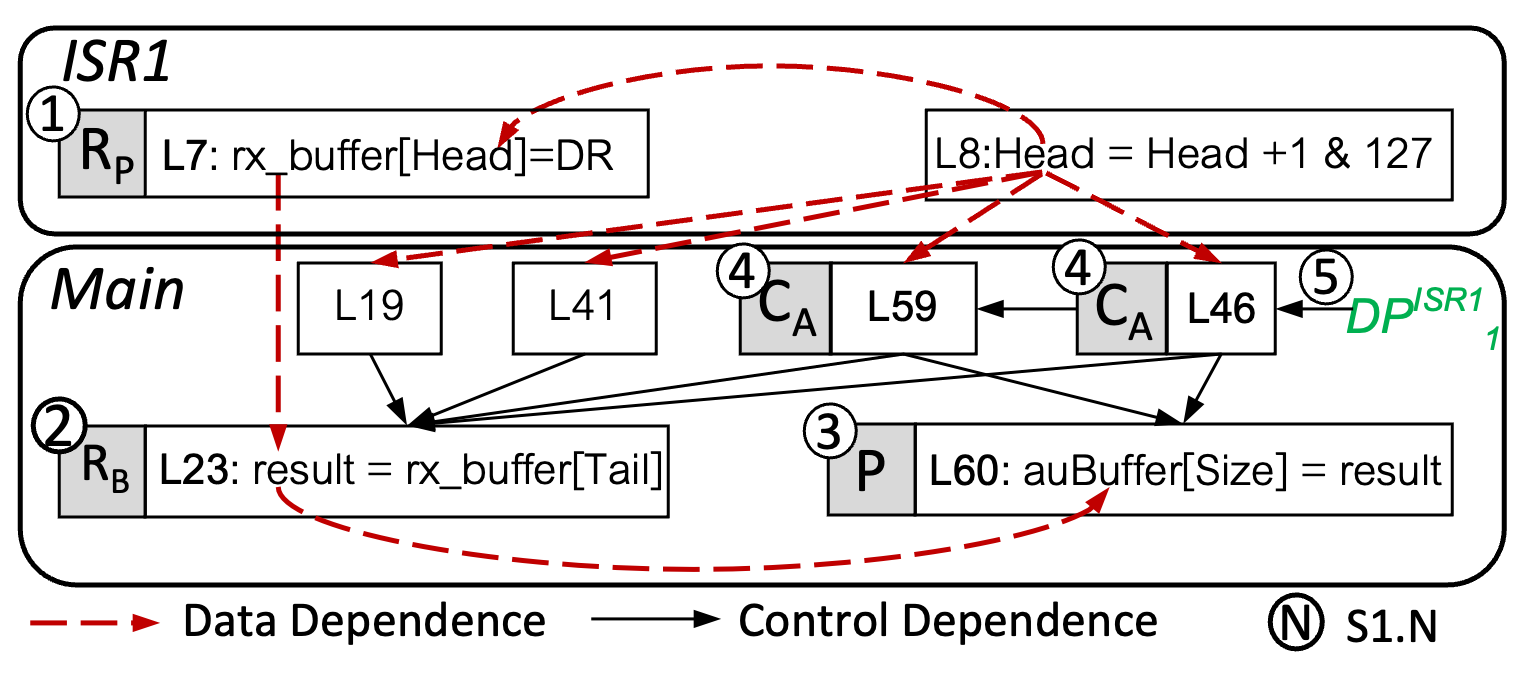}
\caption{CRP input route mapping process with example in~\autoref{lst:heatpress}. (L\#: Line Number in~\autoref{lst:heatpress})}
\label{fig:CRPidentify}
\end{figure}

\section{Design}

\subsection{Objective and Threat Model}
We seek to enhance firmware fuzzing front-end by optimizing the timing and quantity of test case delivery, without altering other components of the existing firmware front and back-end. This is specifically for firmware that receives external data inputs through IRQ mode.
We exclude hardware bugs related to erroneous hardware responses, such as status or control register manipulations, and focus on improving the firmware fuzzer to detect memory-related bugs triggered by external input data, aligning with existing firmware fuzzers.


\subsection{Challenges and High-level Insight}
The diversity and customization of input handing in real firmware implementations pose the following challenges to extract the semantic information for optimal input delivery.

\paragraph{Accurate Availability Check Identification:} While availability checks are performed on global objects modified by ISRs (termed as \textit{availability check variables}), not all these objects are used for such checks. Moreover, global objects used for availability checks may also serve other checks (\eg~\texttt{Head} pointer will also be checked for $\widetilde{R_{B}}$ operations as shown in~\autoref{line:voidread1} in~\autoref{lst:heatpress}), but should not be used for input delivery (avoiding P2). Lastly, availability checks may occur multiple times before $P$ operations; repeated delivery before the same $P$ can lead to the P1 and P3 problems.

\noindent\underline{Insight:} The CRP operations within the same \CRProute exhibit tight data and control dependencies, as illustrated in \autoref{fig:CRPidentify}. Accordingly, we begin by hooking the data I/O access ($R_{P}$), then map and connect other CRP operations to form \CRProutes, allowing us to accurately identify the initial availability check for each \CRProute.

\paragraph{Implicit Length Requirement.} Since input data transfers through multiple buffer objects before processing, identifying the accumulated length in each buffer for length checks is challenging. Tracking the variables indicating length is also tedious. Additionally, the ISR buffer's maximum capacity is determined by firmware logic and differs from the allocated memory space, making static analysis of ISR buffer objects unreliable. Symbol and debug information should not be relied upon as they typically do not exist in stripped firmware.

\noindent\underline{Insight:} Instead of relying on specificity-ignoring applications of existing dynamic and static analysis techniques, we can leverage the inherent characteristics of the input handling programming model to infer the length range: 1) As noted in F3, $P$ operations occur only when length checks (\ie~lower limits) are satisfied; 2) As noted in F1, the ISR uses a ring buffer for input storage, meaning the input is overwritten before retrieval, indicating the input byte length has exceeded the upper limits.
Thus, we propose incrementally increasing the delivery bytes at the delivery point and monitoring the length at which these behaviors occur.

\begin{figure}[t]
\setlength\abovedisplayskip{0pt}
\setlength\belowdisplayskip{0pt}
\centering
\includegraphics[width=\columnwidth]{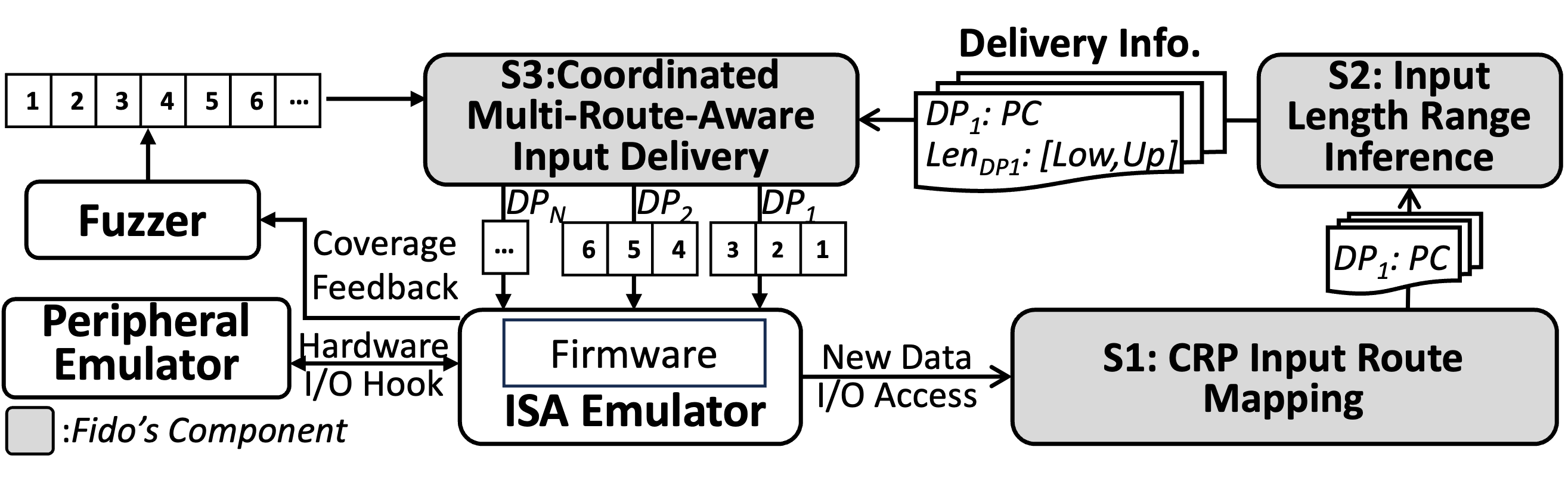}
\caption{Overview of \sys}
\label{fig:arch}
\end{figure}

%% file: tex/4.2-design.tex
\vspace{-1mm}
\subsection{Approach Overview}

\autoref{fig:arch} illustrates the work flow of our solution, \sys, which is activated during firmware fuzzing when peripheral data I/O access occurs. 
\sys first maps the program counter (PC) for each CRP operation in the current \CRProute to pinpoint their delivery points (see~\autoref{sec:CRPidentify}).  
Next, \sys determines delivery length boundaries by adjusting the input length at these points (see ~\autoref{sec:lengthinfer}). The delivery point and its length boundaries constitute the optimal delivery information.

To manage delivery across multiple routes with varying calling contexts, \sys functions as an input delivery extension between the firmware fuzzer's front-end and back-end. In the front-end, it slices and distributes the original test case to each delivery point based on the delivery information (see~\autoref{sec:deliveryalg}). In the back-end, it hooks the PC address of each delivery point. When the firmware hits a delivery point, \sys delivers a fuzzing input slice by triggering the corresponding interrupts.

\subsection{S1: CRP Input Route Mapping}
\label{sec:CRPidentify}
In this subsection, 
we detail the sub-steps for the automatic identification of CRP operations and input routes, as illustrated in \autoref{fig:CRPidentify}.

\paragraph{S1.1: $R_{P}$ Identification.}
During fuzz testing, we hook the MMIO access. If the MMIO access pattern aligns with the DR (as defined in \ppim~\cite{fengp2020p2im}) and the CPU is in handler mode, we confirm the data I/O access ($R_{P}$) is in IRQ mode (\eg~\autoref{line:retrival1}). Alternatively, more precise DR address information can be sourced directly from public MCU reference manuals, as shown by \semu~\cite{zhou2022your}.

\paragraph{S1.2: $R_{B}$ Identification.}
As shown in~\autoref{fig:CRPidentify}, the $R_{P}$ and $R_{B}$ operations are executed in separate firmware contexts. They have no direct control dependence but are directly data-dependent through the ISR buffer (\eg~\texttt{rx\_buffer}). 
As noted in F1, since ISR buffers are global and receive update only from data registers,
we dynamically trace the data flow from the $R_{P}$ operation to determine the final memory address upon ISR exit, identifying the address of the ISR buffer.
By hooking this address, we capture the PC of retrieved buffer data as the $R_{B}$ operations (\eg~\autoref{line:retrival2}) when the firmware execution reaches the hook.




\paragraph{S1.3: $P$ Identification.}
As shown in~\autoref{fig:CRPidentify}, the processing instruction directly depends on $R_{B}$. We use taint analysis from $R_{B}$ to identify the $P$ instructions, setting $R_{B}$ as the source and the $P$ instructions, which are conditional branch instructions involving input (\eg~\texttt{CMP input, R0}), as the sink. 
Note that we only need to identify the nearest $P$ instructions in each \CRProute to distinguish between different \CRProute and pinpoint the availability check operation, as detailed later.
In~\autoref{lst:heatpress}, the \texttt{getRxBuffer} function reads input from \texttt{rx\_buffer} and compares it with a specific number at~\autoref{line:process0}, which is a processing instruction.



\begin{figure}[t]
\setlength\abovedisplayskip{0pt}
\setlength\belowdisplayskip{0pt}
\centering
\includegraphics[width=\columnwidth]{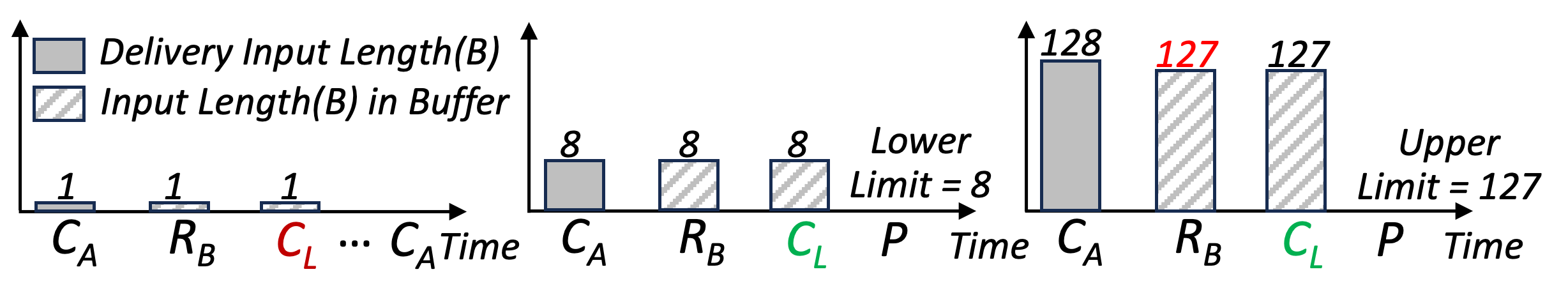}
\caption{Length range inference with example in~\autoref{lst:heatpress}.}
\label{fig:S2}
\end{figure}

\paragraph{S1.4: $C_{A}$ Identification.}
For bare-metal firmware, availability checks often rely on a pointer (\ie~\textit{availability check variable}) indicating the storage location of retrieved data. In RTOS-based firmware, a global semaphore (\ie~\textit{availability check variable}) is also modified after $R_{P}$. During S1.2, we monitor the global memory reading variable, which is written with different values after $R_{P}$ as \textit{availability check variables}. 
These variables are recorded and mapped with corresponding $R_{P}$, $R_{B}$, and $P$ instructions.
We use cross-references of these variables to find conditional branch instructions dependent on them. 
Note that \textit{availability check variables} may be checked before $\widetilde{R_{B}}$ for other purposes. We analyze the control dependence between the check and $P$, considering only checks with reverse control dependence on $P$ and $R_{B}$ operations as real availability checks. As shown in \autoref{fig:CRPidentify}, we identify the \texttt{Head} pointer modified in the ISR function. We find checks in \autoref{line:voidcheck1}, \autoref{line:voidcheck2}, \autoref{line:check1}, and \autoref{line:check2} where branch instructions invoke the \texttt{Head} pointer as shown in~\autoref{fig:CRPidentify}. Only \autoref{line:check1} and \autoref{line:check2}, which have reverse control dependency on the $P$ and $R_{B}$ instructions, are identified as availability checks ($C_{A}$).
Finally, $C_{A}$, $R_{B}$, and $P$ are mapped and connected as a sub-graph in the CFG (\ie~an \CRProute).

\paragraph{S1.5: Delivery Point Identification.}
Multiple availability check points can dominate the same data retrieval point (\eg~\autoref{line:check1} and \autoref{line:check2}) within an \CRProute. To address this, we analyze the control dependencies among these checks and the control flow. We identify the delivery point that has control dependence over others as the initial availability check for each \CRProute. For instance, since the \autoref{line:check1} precedes and controls \autoref{line:check2}, we designate \autoref{line:check1} as a delivery point, mapping it to the corresponding \texttt{UART\_IRQHandler} ISR function, donated as $\textit{DP}^{UART}_{1}$.

\subsection{S2: Input Length Range Inference}
\label{sec:lengthinfer}
We propose a watermark-based method to dynamically determine the limits for input size.
Our method gradually increases the length of the tentatively delivered input through additional interrupts with dummy input
at the identified delivery point, as shown in~\autoref{fig:S2}.
Note that this process needs extra dummy-input but runs only once for each delivery point.


\paragraph{Lower Limit Inference.} We dynamically hook and monitor under what input length the $P$ instructions (identified in S1.3) are reached. 
In the \texttt{Heat\_Press} firmware in \autoref{lst:heatpress}, we start input delivery with triggering UART interrupt at the delivery point (\autoref{line:check1}) and add code hook at $P$ instruction (\autoref{line:process1}). 
We gradually increase the delivery length by triggering ISRs (\texttt{UART\_IRQHandler}) additional times, each adding one byte to the buffer.
Then, we found delivering eight bytes enables the firmware to reach $P$ instructions, setting the lower limit.

\paragraph{Upper Limit Inference.}
We monitor the data written to the ring buffer during ISR. 
Once any old data is overwritten, we note the length and use that as the buffer's upper limit.
In \texttt{Heat\_Press},
we place a memory hook at the beginning of \texttt{rx\_buffer} (identified in S1.2) 
and increase delivery length by triggering additional ISRs, 
pausing the processing function to accumulate input in the buffer. 
When inputs exceed 127 bytes, new data would overwrite the old ones, 
exposing the upper limit of the buffer (on the right of~\autoref{fig:S2}).

%






\subsection{S3: Coordinated Multi-Route-Aware Delivery}
\label{sec:deliveryalg}

\begin{figure}[t]
\setlength\abovedisplayskip{0pt}
\setlength\belowdisplayskip{0pt}
\centering
\includegraphics[width=0.45\columnwidth]{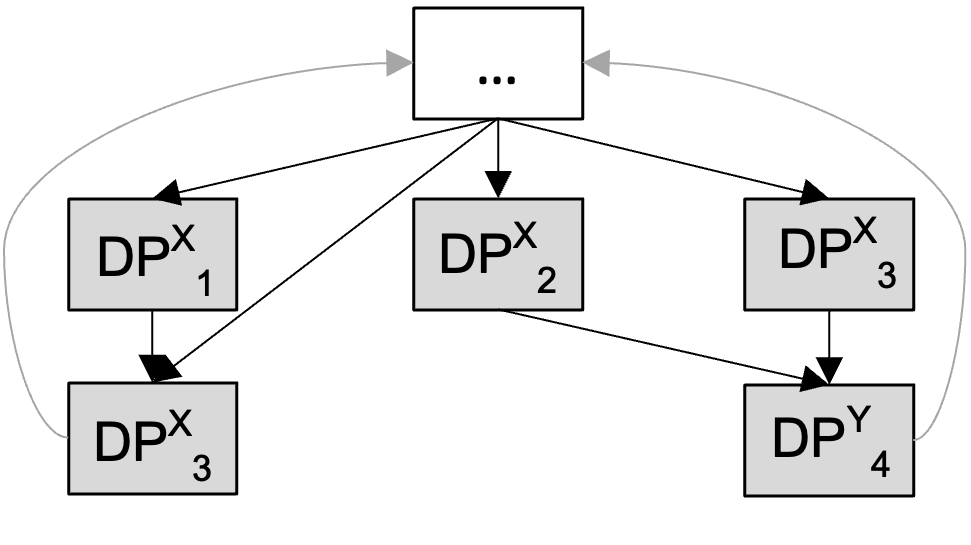}
\caption{The $\sum_{}^{n}\left| CFG(\textit{DP}^{ISR}_{n}) \right|$  = 7 in this figure}
\label{fig:DPcount}
\end{figure}

Firmware can have multiple \CRProutes corresponding to multiple delivery points ($p \in DP$) and each with specific length range requirements ([$LOW_p$, $UP_p$]). 
Different fuzzing rounds may encounter different points at various calling contexts, as mentioned in F4.
Whenever one DP is being handled, one \CRProute is being tested by the fuzzer. In order to test multiple input routes, we propose a coordinated multi-route-aware input delivery algorithm~\autoref{alg:InputSliceAlg}, which intends to partition a given test case ($FI$) into multiple segments: each segment serves one route.

We begin by counting $N$, the total number of already-identified delivery points with distinct calling contexts (different in-edges of the delivery point in the loop CFG), as demonstrated in~\autoref{fig:S2}.
To ensure the fuzzing input can be split into at least $N$ parts (one for each delivery point), each with minimum size $X$ (the smallest lower limit among the current delivery points), we conduct the following steps:
\begin{enumerate}
    \item Line 4-8: Calculate $Len_R = N \times X$, the total length supposed to be reserved for $N$ minimum-sized chunks. If the length of the fuzzing input is shorter than the supposed reservation, manage the trade-off by reducing $N$ until the length of the fuzzing input is no longer shorter. Note that the trade-off here compromises on delivery point coverage.
    \item Line 9: Subtract this reserved length from the total input length. Now, $Len_{FI}$ is the remaining bytes available beyond the reserved chunks.
    \item Line 16-20: For assigning the next $SI$  (segment of the input) at the current point, the algorithm first checks if $Len_{FI}$ plus $X$ meets the current delivery point's minimum length.
    \item Line 21-26: If it does, the algorithm calculates the delivery length ($len_{SI}$) of $SI$ as a random value within current delivery point's limits, extracts (the next) $Len_{SI}$ not-yet-used bytes from $FI$, and delivers $SI$. To maintain fuzzing consistency (\ie~the same fuzzer test-case always produces the same execution results), the algorithm records the delivered $SI$ as the random seed, ensuring reproducibility.
    \item Line 27-32: After allocating $SI$, update the trackers $Len_{FI}$, $Len_{R}$ and $Pos$. Note that $Pos$ ensures that no already-used byte in $FI$ will be reused. 
    \item Line 10-15: When $Len_{R} + Len_{FI}$ is 0, all bytes in $FI$ are delivered, ending the current fuzzing round.
\end{enumerate}

In addition, information of $DP$s is being updated with the \texttt{DPIdentify} function (Steps S1-S2) when new delivery points are identified (Line 12).


\begin{algorithm}[t]
\footnotesize
\caption{Coordinated Multi-Route-Aware Delivery}
\label{alg:InputSliceAlg}
\tcbset{
  colback=white,        
  colframe=black,       
  boxrule=0.5pt,        
  sharp corners,        
  width=\columnwidth,     
  left=0pt, right=0pt,  
  top=0pt, bottom=0pt,  
}
\begin{tcolorbox}
\textbf{Input:} Entry Point ($EP$), Fuzzing Input($FI$), Delivery Point ($DP$)
\end{tcolorbox}
\begin{algorithmic}[1]
\State{$Pos = 0$, $PC = EP$}
\State{$N=\sum_{}^{n}\left| CFG(\textit{DP}_{n}) \right|$}
\State{$X = \min_{p \in DP} LOW_p$}
\State{$Len_{R} = N*X$}
\While{$Len_{FI} < Len_{R}$}
    \State{$N = N-1$};
    \State{$Len_{R} = N*X$};
\EndWhile
\State{$Len_{FI} -= Len_{R}$}
\While{true}
    \State {$PC=Execute(PC)$};
    \If{$PC == p \in DP$}
        \If{$Len_{FI} + Len_{R} == 0$}
            \State \textbf{return};
        \Else
            \State {$\Delta = Len_{FI}+X-LOW_p$};
            \If{$\Delta \le 0$}
                \State{$Len_{SI}=Len_{FI}+X$};
                \State{$SI= FI_{[Pos,Pos+Len_{FI}+X]} \parallel \underbrace{00 \ldots 0}_{|\Delta|}$};
                \State{$Delivery(SI)$};
            \Else 
                \State{$t=\min\left(Len_{FI}+X, UP_p\right)-LOW_p$};
                \State{$Len_{SI}=Rand(FI_{[Pos]})\mod t +LOW_p$};
                \State{$SI = FI_{[Pos,Pos+Len_{SI}]}$};
                \State{$Delivery(SI)$};
            \EndIf
            \State{$Len_{FI} -= Len_{SI} - X$};
            \State{$Len_{R} -= X$};
            \State{$Pos=Pos + Len_{SI}$};
            \If{$Len_{R} == 0$} \State{$X = 0$};\EndIf
        \EndIf
    \EndIf
    \If{$PC == New \ R_P \ or \ R_B$}
        \State{$DP += DPIdentify(PC)$} \label{line:S1-S3}
    \EndIf
\EndWhile
\end{algorithmic}
\end{algorithm}


%% file: tex/5-impl.tex
\vspace{-1mm}
\section{Implementation}
\label{sec:imp}
We implemented our prototype, \sys, as a plugin for three state-of-the-art firmware fuzzers: \mulitfuzz, \semu, and \fuzzware. 
These fuzzers originally employ various delivery strategies as listed in \autoref{tab:fuzzdelivery}
with single- or multi-stream support.
This plugin implementation demonstrates \sys's compatibility and allows for a comprehensive evaluation of different delivery strategies.

\sys consists of a set of static and dynamic analysis modules that interact via a Ghidra server~\cite{ghidra-server}. The static analysis module employs Ghidra~\cite{ghidra} for the sub-steps of S1. In step S1.3, \sys leverages Ghidra's \texttt{getDescendants} for data flow taint analysis to find the nearest compare branch instruction (PCode type \texttt{INT\_EQUAL}) as $P$ instructions. In step S1.4, it uses Ghidra's \texttt{getReferencesTo} API to collect reference information for global variables. Step S1.5 involves using Ghidra's \texttt{getSources} API to backtrack the control flow of each basic block to identify initial availability check instructions.

The dynamic analysis for steps S1, S2, and S3 is customized for each emulator
back-end with a similar interface: \fuzzware and \semu utilize Unicorn, while
\mulitfuzz is based on Icicle~\cite{chesser2023icicle}. We use Unicorn as an
example to illustrate our implementation. The same idea applies to Icicle,
albeit with different APIs. For Unicorn, in step S1.1, we use the
\texttt{UC\_HOOK\_MEM\_READ} API for MMIO access monitoring to identify $R_P$.
In step S1.2, we use \texttt{UC\_HOOK\_MEM\_WRITE} API to track the global ISR
buffer access and identify modified global variables. Static analysis in steps
S1.3, S1.4, and S1.5 may struggle with indirect jumps, which we address using a
hybrid method similar to recent works~\cite{zhu2021constructing,
chesser2024ffxe} to solve them on demand.
In step S2, \texttt{UC\_HOOK\_CODE} API at each $P$ instruction determines the
lower limit, while \texttt{UC\_HOOK\_MEM\_WRITE} API at the beginning memory
address of ISR buffer determines the upper limit. We note that dynamic hooks in
S1 and S2 are a one-time setup and are removed once delivery points and lengths
are determined. In step S3, we use the \texttt{UC\_HOOK\_CODE} API to hook all
delivery points ($DPs$) to manage the delivery of fuzzing input generated by the
fuzzer front-end, as outlined in Algorithm~\ref{alg:InputSliceAlg}.
Specifically, we trigger the corresponding interrupt for $SI$ delivery at
specific times at the $DP$ Hook.

%% file: tex/6-eval.tex
\vspace{-1mm}
\section{Evaluation}


We evaluated \sys to answer the following research questions (\textbf{RQs}):

\noindent \textbf{RQ1:} Can \sys automatically identify \CRProutes in both unit
tests and real-world firmware? (\autoref{sec:eval:info})

\noindent \textbf{RQ2:} By addressing the delivery issues outlined
in~\autoref{sec:problems}, to what extent does \sys improve code coverage and
bug-finding capability compared to the ad hoc delivery strategies used by SOTA
firmware fuzzers? (\autoref{sec:eval:fuzz})

\noindent \textbf{RQ3:} How does \sys compare with other interrupt-driven
firmware fuzzers, such as \aid, in addressing the identified delivery issues?
(\autoref{sec:eval:aid})

\noindent \textbf{RQ4:} To what extent does each component (\ie~S1, S2, and S3)
contribute to fuzzing effectiveness? (\autoref{sec:eval:ablation})

\paragraph{Firmware Samples Collection.} 
We selected 28 unit test samples from our large-scale empirical study in~\autoref{sec:programmodel}. These samples include driver code for common data I/O peripherals (GPIO, UART, I2C, ADC), HALs from top MCU vendors (STM32, NXP, Arduino), and RTOS kernels (RIoT, Nuttx).
We tested 25 real-world firmware samples, including 22 that were tested by \semu, \fuzzware, \aid and \mulitfuzz, and have at least one data peripheral in IRQ mode. 
We excluded samples without peripherals operating in IRQ mode.
We also excluded samples that only timer peripherals in interrupt mode, as \sys does not contribute to them. 
We also excluded DMA firmware and \texttt{STM\_PLC} samples, as neither \aid nor \sys supports DMA emulation and nested interrupt triggering.
We also incorporated two BLE GATT server examples from the MbedOS 
BLE project~\cite{Mbedblerepo} (\texttt{Gatt\_Clientupdate} and \texttt{Gatt\_Serverupdate}) 
and one \texttt{BLE\_HCI} example from the Zephyr project~\cite{Zephyrblerepo}.
Details about these samples, including MCU model, OS/library, and total basic block number, 
are detailed in~\autoref{tab:detailsample} in Appendix~\ref{app:detailinfo}.

\paragraph{Experiment Setup.}
All experiments were conducted on a PC equipped with an Intel Xeon Platinum
8350C processor at 2.60GHz, 256GB RAM, and a 960GB SSD storage.

\begin{table*}[t]
\caption{Code coverage (median of 5 trials after 24-hours) using \sys compared to original delivery methods. Shaded areas means indicating additional coverage from bug exploits. Problem (P) denotes delivery problems under RR or fuzz mode. Growth Rate (GR) for RR and fuzz modes with significant changes are marked in bold (based on a Mann-Whitney U test with a 0.05 significance threshold).}
\centering
\begin{adjustbox}{max width=\textwidth}
\begin{tabular}{l|l|ccccccc|ccccccc}
\hline
\multicolumn{1}{l|}{}                                    & \multicolumn{1}{l|}{}                                   & \multicolumn{7}{c|}{\textbf{\fuzzware}}                                                                                                                                                                & \multicolumn{7}{c}{\textbf{\textit{MultiFuzz}}}                                                                                                                                                               \\
\multicolumn{1}{l|}{\multirow{-2}{*}{\textbf{Firmware}}} & \multicolumn{1}{l|}{\multirow{-2}{*}{\textbf{Feature}}} & \textbf{RR}    & \textbf{Fuzz}  & \textbf{w.\sys}                        & \textbf{P(RR)} & \textbf{GR(RR)}                           & \textbf{P(Fuzz)} & \multicolumn{1}{c|}{\textbf{GR(Fuzz)}}    & \textbf{RR}    & \textbf{Fuzz} & \textbf{w.\sys}                        & \textbf{P(RR)} & \textbf{GR(RR)}                           & \textbf{P(Fuzz)} & \textbf{GR(Fuzz)}                         \\ \hline
3DPrinter                                                & F1,F2,F3,F4                                             & 786            & 780            & \textbf{931}                           & P2-P4            & \textbf{+18.4\%}                          & P2-P4            & \textbf{+19.3\%}                          & 4,193          & 3,642         & \textbf{4,411}                         & P2-P4            & +5.2\%                                    & P2-P4            & \textbf{+21.1\%}                          \\
Bootstrap(SPI)                                           & F1,F3,F4                                                & 956            & 950            & \textbf{998}                           & P3               & +4.4\%                                    & P3               & +5.1\%                                    & 982            & 1,184         & \textbf{1,198}                         & P3               & \textbf{+22.0\%}                          & P3               & +1.2\%                                    \\
Bootstrap(UART)                                          & F1,F3,F4                                                & 994            & 951            & \textbf{1,878}                         & P1,P3-P4         & \textbf{+88.9\%}                          & P1,P3-P4         & \textbf{+97.5\%}                          & 1,289          & 1,986         & \textbf{1,986}                         & P1,P3-P4         & \textbf{+54.1\%}                          & P1,P3-P4         & +0.0\%                                    \\
CCN-Lite-Relay                                           & F1,F3,F4                                                & 491            & 556            & \textbf{1,054}                         & P3-P4            & \textbf{+114.7\%}                         & P3-P4            & \textbf{+89.6\%}                          & 4,077          & 4,445         & \textbf{4,472}                         & P3-P4            & \textbf{+9.7\%}                           & P3               & +0.6\%                                    \\
$\mu$tasker\_USB                                         & F1,F3,F4                                                & 1,269          & 1,253          & \textbf{1,518}                         & P2-P4            & \textbf{+19.6\%}                          & P2-P3            & \textbf{+21.1\%}                          & 1,995          & 1,924         & \textbf{2,129}                         & P2-P4            & +6.7\%                                    & P2-P3            & +10.7\%                                   \\
Console                                                  & F1,F2,F3                                                & 711            & 712            & \textbf{794}                           & P2-P3            & +11.7\%                                   & P2-P3            & +11.5\%                                   & 1,165          & 1,161         & \textbf{1,171}                         & P2-P3            & +0.5\%                           & P2-P3            & \textbf{+0.9\%}                           \\
Echo\_Server                                             & F3& 2,854          & 2,852          & \textbf{2,905}                         & P3               & +1.8\%                                    & P3               & +1.9\%                                    & 3,553          & 3,567         & \textbf{3,569}                         & P3               & +0.5\%                           & P3               & +0.1\%                                    \\
Gateway                                                  & F1,F3,F4                                                & 2,362          & 2,712          & \textbf{2,756}                         & P3-P4            & \textbf{+16.7\%}                          & P1-P4            & +1.6\%& 2,968          & 2,882         & \textbf{3,188}                         & P3-P4            & \textbf{+7.4\%}                           & P1-P4            & \textbf{+10.6\%}                          \\
Gnrc\_networking                                          & F1,F3,F4                                                & 421            & 416            & \textbf{668}                           & P3-P4            & \textbf{+58.7\%}                          & P3-P4            & \textbf{+60.6\%}                          & 1,849          & 1,779         & \textbf{2,136}                         & P3-P4            & \textbf{+15.5\%}                          & P3-P4            & \textbf{+20.1\%}                          \\
GPSTracker                                               & F1,F2,F3,F4                                             & 661& 977            & \textbf{1,011}                         & P2-P4            & \textbf{+53.0\%}& P2-P4            & +3.5\%                                    & 1,227 & 1440          & \textbf{1,589}                         & P2-P4            & \textbf{+29.5\%}                          & P2-P4            & \textbf{+10.3\%}                          \\
Heat\_Press                                              & F1,F2,F3,F4                                             & 551            & 555            & \textbf{570}& P2-P4            & \textbf{+3.4\%}& P2-P4            & \textbf{+2.7\%}& 573            & 580           & \textbf{601}                           & P2-P4            & +4.9\%                                    & P2-P4            & +3.6\%                                    \\
L2cap\_Processor                                         & F3& \textbf{1,001} & 1,001          & \textbf{1,001}                         & P3               & +0.0\%                           & P3               & +0.0\%                           & 1,002          & 1,021         & \textbf{1,170}                         & P3               & \textbf{+16.8\%}                          & P3               & \textbf{+14.6\%}                          \\
LiteOS\_IoT                                              & F1,F3,F4                                                & 738& 746            & \textbf{1,333}                         & P3-P4            & \textbf{+80.6\%}& P1-P4            & \textbf{+78.6\%}                          & 1,375          & 1,380         & \textbf{1,377}                         & P3-P4            & \textbf{+0.1\%}                           & P1-P4            & -0.2\%                                   \\
PLC                                                      & F1,F2,F3                                                & 638            & 640   & 642 & P2-P3            & +0.3\% & P2-P3            & +0.3\% & 640            & 640           & \cellcolor[HTML]{DEE0E3}\textbf{1,838} & P2-P3            & \cellcolor[HTML]{DEE0E3}\textbf{+187.2\%} & P2-P3            & \cellcolor[HTML]{DEE0E3}\textbf{+187.2\%} \\
Filesystem                                               & F1,F3,F4& \textbf{-}     & \textbf{-}     & \textbf{-}                             & -                & -                                         & -                & -                                         & 1,374          & 1,352         & \textbf{1,414}                         & P3               & +2.9\%                                    & P3               & +4.6\%                                    \\
Snmp\_Server                                             & F3& 1,032          & 1,032 & \textbf{1,045}                         & P3               & +1.3\%                                    & P3               & +1.3\%                                    & 1,066          & 1,083         & \textbf{1,297}                         & P3               & \textbf{+21.7\%}                          & P3               & \textbf{+19.8\%}                          \\
Soldering\_Iron                                          & F1,F3,F4                                                & 2,177          & \textbf{2,280} & 2,267                                  & P3               & +4.1\%                           & P3               & -0.6\%                           & 2,675          & 2,799         & \textbf{3,271}                         & P3               & \textbf{+22.3\%}                          & P3               & \textbf{+16.9\%}                          \\
Steering\_Control                                        & F1,F3,F4                                                & 587            & 594            & \textbf{606}                           & P3               & +3.1\%                          & P3               & +1.9\%                           & 652            & 655           & \textbf{660}                           & P3               & +1.2\%                           & P3               & +0.8\%                                    \\
Zephyr\_SocketCan                                         & F1,F3                                                   & 2,583          & \textbf{2,662} & 2,660                                  & P3               & +3.0\%                                    & P3               & -0.1\%                                    & 3,334          & 2,880         & \textbf{3,341}                         & P3               & +0.2\%                           & P3               & \textbf{+16.0\%}                          \\ \hline
Client-Gattupdate                                        & F1,F3,F4                                                & -              & -              & -                                      & -                & -                                         & -                & -                                         & 4,333          & 3,051         & \cellcolor[HTML]{DEE0E3}\textbf{7,454} & P3-P4            & \cellcolor[HTML]{DEE0E3}\textbf{+72.0\%}  & P3-P4            & \cellcolor[HTML]{DEE0E3}\textbf{+144.3\%} \\
Server-Gattupdate                                        & F1,F3,F4                                                & -              & -              & -                                      & -                & -                                         & -                & -                                         & 10,117         & 10,846        & \textbf{11,018}                        & P3-P4            & +8.9\%                                    & P3               & +1.6\%                                    \\
BLE-HCI                                                  & F1,F3,F4                                                & 2,523          & 2,576          & \textbf{2,784}                         & P3-P4            & +10.3\%                                   & P3-P4            & \textbf{+8.1\%}                                    & -              & -             & -                                      & -                & -                                         & -                & -                                         \\ \hline
\end{tabular}
\end{adjustbox}
\label{tab:eval:med-cov}
\caption*{%
    \scriptsize
    \raggedright
    -: The original firmware fuzzer fails to reach the stage where firmware receives external input (\eg~getting stuck in initialization stage after 24 hours).
}
\end{table*}

\subsection{Delivery Information Identification (RQ1)}
\label{sec:eval:info}

A unit-test samples usually tests a single peripheral in an infinite main loop.
We manually verify the delivery information identified by \sys once a main loop
completes. In contrast, real-world firmware often contains multiple \CRProutes
with varying activation conditions. To cover as many \CRProutes as possible, we
collect the resulting delivery information after 24 hours of fuzzing and then
manually verify its accuracy.




\sys accurately identifies delivery information of each \CRProute with details listed in Appendix~\ref{app:dpinfo} and Appendix~\ref{app:detailinfo}. 
Firmware varies in the number of \CRProutes, peripherals, and length requirements. 
Within a firmware sample, some \CRProutes consistently serve as main inputs during fuzzing, 
while others appear only under specific conditions. 
This variability highlights the challenges in automatically and accurately identifying the different stages within an \CRProute. 



\paragraph{Time Usage.} We measure the time from hitting the retrieval hooks to
extracting the delivery information. Identifying one delivery point (S1) takes
3--10 seconds (8.84 seconds on average), while length inference (S2) takes 2--7
seconds (5.06 seconds on average). Overall, extracting delivery information per
sample takes less than 20 seconds, highlighting the efficiency of our approach.
Additionally, for an \CRProute, delivery information extraction is a one-time
effort.



\subsection{Fuzzing Efficiency Improvement (RQ2)}
\label{sec:eval:fuzz}

For RQ2, we compare the original delivery methods (RR and Fuzz modes for both
interrupt selection and interval) shipped with SOTA firmware fuzzer
(\ie~\fuzzware, \mulitfuzz and \semu), against the \sys method integrated with
the same fuzzing tools. For each target, we conducted five iterations using
different random seeds.


The fuzzing improvement potential of \sys depends on the extent to which the existing ad-hoc delivery strategies suffer from problems P1-P4.
Addressing P1 and P2 maximizes fuzzing input capacity, solving P3 allows the executor to focus on exploring input processing code spaces where real exploitable bugs exist faster, and resolving P4 enables fuzzing to achieve greater code coverage.
For the samples \ie~\texttt{6LoWPAN\_Receiver}, \texttt{6LoWPAN\_Transmitter}, \texttt{P2IM\_Drone},
\sys shows minimal performance improvement, as detailed in Appendix~\ref{app:detailinfo}.
A manual check reveals that the fuzzer rarely triggered data peripherals in interrupt mode for these samples. For instance, the UART peripheral of \texttt{P2IM\_Drone} in interrupt mode is only used during the firmware setup process.


\subsubsection{Code Coverage Improvement}
\label{sec:eval:coverage}

As shown in \autoref{tab:eval:med-cov}, \sys achieves higher median coverage across most samples than \mulitfuzz and \fuzzware, regardless of RR or fuzz mode, particularly for complex firmware with multiple \CRProutes (F4) like \texttt{Gateway}, \texttt{LiteOS\_IoT}, \texttt{3D\_printer}, and \texttt{CCN-Lite-Relay}.

This is because RR or fuzz mode configurations struggle to deliver inputs promptly, particularly for routes that are activated only under specific conditions.
For instance, the \texttt{Gateway} firmware primarily uses UART for continuous command processing, while other routes are only activated under specific command value via UART.
When the command value equals \texttt{0x78}, the I2C peripheral can be enabled, hitting the I2C availability check. If the check fails, this input route cannot be accessed again in the same fuzzing round. In both RR and Fuzz modes, delivering input at this precise moment is extremely difficult. By hooking the I2C availability check operation, we found that in \mulitfuzz's RR delivery method, although the I2C availability check was reached 790,060 times, it failed to trigger the I2C interrupt for input delivery, missing coverage of I2C input processing.

Addressing P2-P3 accelerates coverage exploration. 
As also shown in~\autoref{fig:cov} in Appendix~\ref{app:cov}, \sys increases coverage faster than the RR or fuzz strategy.
Note that in RR or
fuzz delivery strategies, each byte is delivered after certain
number of basic blocks are executed, allowing enough time for inputs
to be read, so the overfeeding problem P1 is less common,
except for input routes containing stricter length checks like \texttt{BLE\_Bootstrap}

\begin{figure}[t]
\setlength\abovedisplayskip{0pt}
\setlength\belowdisplayskip{0pt}
\centering
\includegraphics[width=\columnwidth]{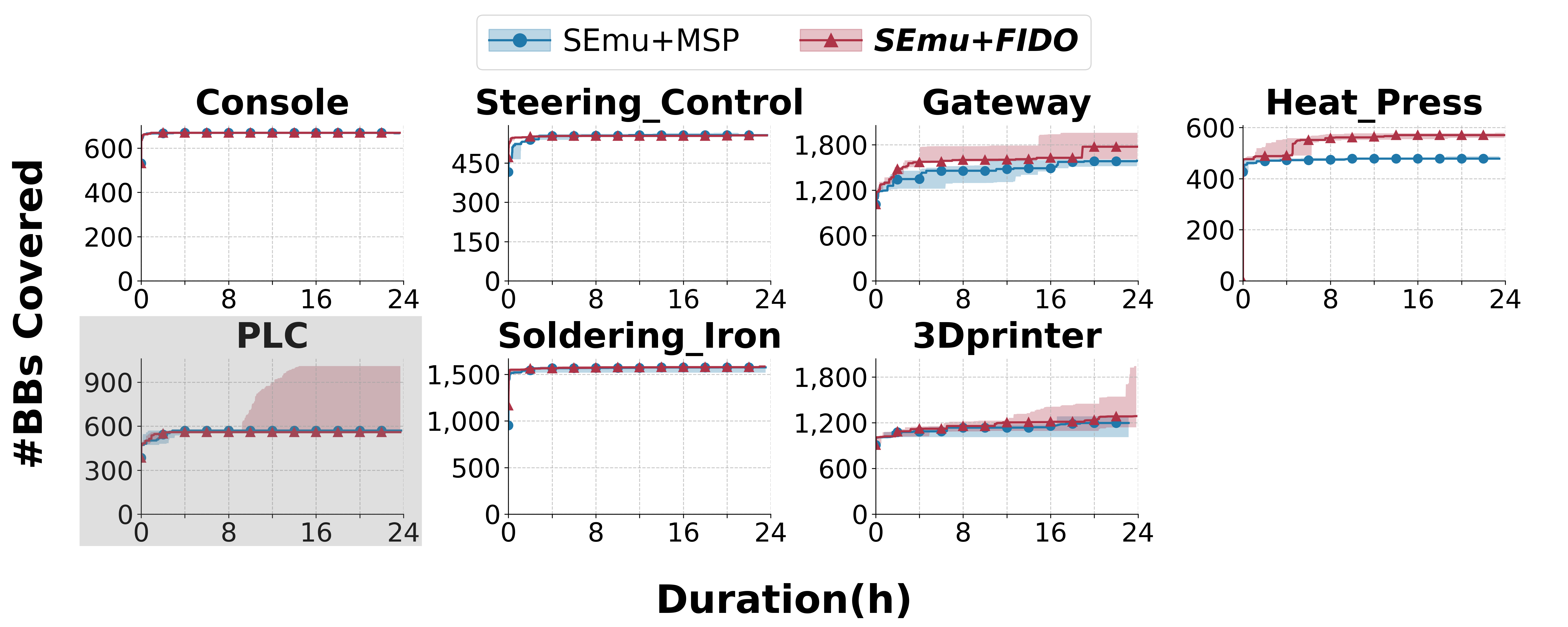}
\caption{Code coverage comparison between {\semu}+{MSP} and {\semu}+{\sys} (only
7 samples are shown, as the other samples require peripheral models that
\semu does not extract).}
\label{fig:semucov}
\end{figure}

As illustrated in~\autoref{fig:semucov}, \sys achieves greater coverage than \semu, which relies on manually specified delivery points. While \semu can address P2 and P3 if the delivery points match those identified by \sys, as seen with \texttt{Console}, it cannot infer quantity requirements. This limitation results in delivering all input at once without length restrictions, leading to the P1 problem across all test samples. Additionally, for firmware with multiple \CRProutes like \texttt{Heat\_Press}, \texttt{Gateway}, and \texttt{Steering\_Control}, \semu supports input delivery at only a single point, leaving P4 unresolved.

\begin{table}[t]
\caption{Crash detection comparison with and without \sys over 48 hours: Crash count represents the total unique crashes, excluding false positives. Time is denoted as (hh:mm:ss). Newly discovered vulnerability is in green bold text. Fac. shows the reduction factor in time when using \sys for bug discovery compared to the original method and have significant changes are marked in bold (based on a Mann-Whitney U test with a 0.05 significance threshold).}
\centering
\begin{adjustbox}{width=\columnwidth}
\begin{tabular}{clrrccr}
\hline
\multicolumn{1}{c|}{\textbf{Firmware}}                & \multicolumn{1}{c|}{\textbf{Bug[Report]}}                                      & \multicolumn{2}{c|}{\textbf{\# Crash Count}}                                  & \multicolumn{3}{c}{\textbf{Minium Discovery Time}}                                                                                                  \\
\multicolumn{1}{c|}{}                                 & \multicolumn{1}{c|}{\textbf{Type-Func./CVE-20..}}                             & \multicolumn{1}{c}{\textbf{Original}} & \multicolumn{1}{c|}{\textbf{w.\sys}} & \textbf{Original}                                      & \textbf{w.\sys}                                        & \multicolumn{1}{c}{\textbf{Fac.}} \\ \hline
\multicolumn{7}{c}{\textbf{\semu(Original = MSP)}}                                                                                                                                                                                                                                                                                                                          \\ \hline
\multicolumn{1}{c|}{Heat\_Press}                      & \multicolumn{1}{l|}{OOB-FC3\cite{HeatPressdescrip}}                            & {\color[HTML]{CB0000} Fail}           & \multicolumn{1}{r|}{\textbf{175}}    & -                                                      & \textcolor{lightgray}{00:}09:31                        & \textbf{\textgreater302}          \\ \hline
\multicolumn{1}{c|}{}                                 & \multicolumn{1}{l|}{OOB-FC1\cite{PLC1descrip}}                                 & 24                                    & \multicolumn{1}{r|}{\textbf{112}}    & 03:23:18                                               & \textcolor{lightgray}{00:}44:24                        & \textbf{4.58}                     \\
\multicolumn{1}{c|}{}                                 & \multicolumn{1}{l|}{OOB-FC3\cite{PLC2descrip}}                                 & {\color[HTML]{CB0000} Fail}           & \multicolumn{1}{r|}{\textbf{17}}     & -                                                      & 01:56:40                                               & \textbf{\textgreater24}           \\
\multicolumn{1}{c|}{}                                 & \multicolumn{1}{l|}{OOB-FC15\cite{PLC3descrip}}                                & 135                                   & \multicolumn{1}{r|}{\textbf{175}}    & \textcolor{lightgray}{00:}14:23                        & \textcolor{lightgray}{00:}08:52                        & \textbf{1.62}                     \\
\multicolumn{1}{c|}{\multirow{-4}{*}{PLC}}            & \multicolumn{1}{l|}{OOB-FC16\cite{PLC4descrip}}                                & 100                                   & \multicolumn{1}{r|}{\textbf{185}}    & 01:53:08                                               & \textcolor{lightgray}{00:}33:48                        & \textbf{3.35}                     \\ \hline
\multicolumn{7}{c}{\textbf{\fuzzware(Original = The better results of RR and Fuzz)}}                                                                                                                                                                                                                                                                                        \\ \hline
\multicolumn{1}{c|}{Heat\_Press}                      & \multicolumn{1}{l|}{OOB-FC3\cite{HeatPressdescrip}}                            & 1236                                  & \multicolumn{1}{r|}{\textbf{1765}}   & \textcolor{lightgray}{00:}56:12                        & \textcolor{lightgray}{00:}19:52                        & \textbf{2.83}                     \\ \hline
\multicolumn{1}{c|}{}                                 & \multicolumn{1}{l|}{OOB-FC1\cite{PLC1descrip}}                                 & 258                                   & \multicolumn{1}{r|}{\textbf{399}}    & \textcolor{lightgray}{00:}42:35                        & \textcolor{lightgray}{00:}46:27                        & 0.92                              \\
\multicolumn{1}{c|}{}                                 & \multicolumn{1}{l|}{OOB-FC3\cite{PLC2descrip}}                                 & 68                                    & \multicolumn{1}{r|}{\textbf{129}}    & 05:07:08                                               & 04:48:25                                               & 1.06                              \\
\multicolumn{1}{c|}{}                                 & \multicolumn{1}{l|}{OOB-FC15\cite{PLC3descrip}}                                & 244                                   & \multicolumn{1}{r|}{\textbf{385}}    & \textcolor{lightgray}{00:}10:23                        & \textcolor{lightgray}{00:}09:56                        & 1.05                              \\
\multicolumn{1}{c|}{\multirow{-4}{*}{PLC}}            & \multicolumn{1}{l|}{OOB-FC16\cite{PLC4descrip}}                                & 88                                    & \multicolumn{1}{r|}{\textbf{1329}}   & 01:28:57                                               & \textcolor{lightgray}{00:}21:26                        & \textbf{4.15}                     \\ \hline
\multicolumn{1}{c|}{}                                 & \multicolumn{1}{l|}{21-3319\cite{3319descrip}}                                 & 1590                                  & \multicolumn{1}{r|}{\textbf{1620}}   & \textcolor{lightgray}{00:}27:34                        & \textcolor{lightgray}{00:}25:13                        & 1.09                              \\
\multicolumn{1}{c|}{}                                 & \multicolumn{1}{l|}{21-3320\cite{3320descrip}}                                 & 1                                     & \multicolumn{1}{r|}{\textbf{18}}     & 02:33:05                                               & 03:42:53                                               & 0.69                              \\
\multicolumn{1}{c|}{\multirow{-3}{*}{Echo\_Server}}   & \multicolumn{1}{l|}{20-10064\cite{10064descrip}}                               & 2                                     & \multicolumn{1}{r|}{\textbf{8}}      & 10:34:24                                               & 11:30:48                                               & 0.92                              \\ \hline
\multicolumn{1}{c|}{Bootstrap(UART)}                  & \multicolumn{1}{l|}{21-3329\cite{3329descrip}}                                 & {\color[HTML]{CB0000} Fail}           & \multicolumn{1}{r|}{\textbf{1}}      & -                                                      & 35:26:12                                               & \textgreater1.41                  \\ \hline
\multicolumn{1}{c|}{}                                 & \multicolumn{1}{l|}{20-10065\cite{10065descrip}}                               & 66                                    & \multicolumn{1}{r|}{\textbf{774}}    & 07:31:43                                               & \textcolor{lightgray}{00:}07:02                        & \textbf{64.23}                    \\
\multicolumn{1}{c|}{\multirow{-2}{*}{Bootstrap(SPI)}} & \multicolumn{1}{l|}{20-10066\cite{10066descrip}}                               & 15                                    & \multicolumn{1}{r|}{\textbf{225}}    & 05:18:11                                               & 03:45:45                                               & \textbf{1.41}                     \\ \hline
\multicolumn{1}{c|}{BLE-HCI}                          & \multicolumn{1}{l|}{{\color[HTML]{009901} \textbf{NPD-ull\_conn.}}}            & {\color[HTML]{CB0000} Fail}           & \multicolumn{1}{r|}{\textbf{1}}      & -                                                      & 08:16:22                                               & \textbf{\textgreater6.04}         \\ \hline
\multicolumn{7}{c}{\textbf{\textit{MultiFuzz}(Original = The better results of RR and Fuzz)}}                                                                                                                                                                                                                                                                                       \\ \hline
\multicolumn{1}{c|}{Heat\_Press}                      & \multicolumn{1}{l|}{OOB-FC3\cite{HeatPressdescrip}}                            & 127                                   & \multicolumn{1}{r|}{\textbf{302}}    & \textcolor{lightgray}{00:}04:02                        & \textcolor{lightgray}{00:}01:12                        & \textbf{3.36}                     \\ \hline
\multicolumn{1}{c|}{}                                 & \multicolumn{1}{l|}{OOB-FC1\cite{PLC1descrip}}                                 & 91                                    & \multicolumn{1}{r|}{\textbf{374}}    & \textcolor{lightgray}{00:}\textcolor{lightgray}{00:}41 & \textcolor{lightgray}{00:}\textcolor{lightgray}{00:}20 & \textbf{2.05}                     \\
\multicolumn{1}{c|}{}                                 & \multicolumn{1}{l|}{OOB-FC3\cite{PLC2descrip}}                                 & 7                                     & \multicolumn{1}{r|}{\textbf{15}}     & 04:44:55                                               & 04:31:55                                               & 1.05                              \\
\multicolumn{1}{c|}{}                                 & \multicolumn{1}{l|}{OOB-FC15\cite{PLC3descrip}}                                & 6                                     & \multicolumn{1}{r|}{\textbf{9}}      & \textcolor{lightgray}{00:}26:22                        & \textcolor{lightgray}{00:}02:34                        & \textbf{10.27}                    \\
\multicolumn{1}{c|}{\multirow{-4}{*}{PLC}}            & \multicolumn{1}{l|}{OOB-FC16\cite{PLC4descrip}}                                & 6                                     & \multicolumn{1}{r|}{\textbf{20}}     & \textcolor{lightgray}{00:}28:17                        & \textcolor{lightgray}{00:}26:24                        & 1.07                              \\ \hline
\multicolumn{1}{c|}{}                                 & \multicolumn{1}{l|}{NPD-strtok\_xx\cite{GPSTracker1descrip}}              & 5                                     & \multicolumn{1}{r|}{\textbf{5}}      & 02:01:33                                               & 01:28:02                                               & 1.38                              \\
\multicolumn{1}{c|}{}                                 & \multicolumn{1}{l|}{NPD-strstr\_xx\cite{GPSTracker2descrip}}           & {\color[HTML]{CB0000} Fail}           & \multicolumn{1}{r|}{\textbf{3}}      & -                                                      & 05:01:49                                               & \textbf{\textgreater9.6}                     \\
\multicolumn{1}{c|}{\multirow{-3}{*}{GPSTracker}}     & \multicolumn{1}{l|}{NPD-strstr\_xx\cite{GPSTracker3descrip}}            & 2                                     & \multicolumn{1}{r|}{\textbf{3}}      & 09:51:50                                               & 05:45:03                                               & 1.72                              \\ \hline
\multicolumn{1}{c|}{}                                 & \multicolumn{1}{l|}{ING-Sysex.\cite{Gateway2descrip}}                           & 6                                     & \multicolumn{1}{r|}{\textbf{8}}      & \textcolor{lightgray}{00:}03:55                        & \textcolor{lightgray}{00:}05:46                        & 0.68                              \\
\multicolumn{1}{c|}{}                                 & \multicolumn{1}{l|}{OOB-setPin.\cite{Gateway3descrip}}                     & \textbf{30}                           & \multicolumn{1}{r|}{20}              & \textcolor{lightgray}{00:}02:39                        & \textcolor{lightgray}{00:}01:51                        & 1.43                              \\
\multicolumn{1}{c|}{}                                 & \multicolumn{1}{l|}{UPD-TxCplt\cite{Gateway4descrip}}                  & \textbf{10}                           & \multicolumn{1}{r|}{8}               & \textcolor{lightgray}{00:}\textcolor{lightgray}{00:}01 & \textcolor{lightgray}{00:}\textcolor{lightgray}{00:}01 & 1                                 \\
\multicolumn{1}{c|}{}                                 & \multicolumn{1}{l|}{NPD-pwm\_start\cite{Gateway6descrip}}                      & 97                                    & \multicolumn{1}{r|}{\textbf{764}}    & \textcolor{lightgray}{00:}17:20                        & \textcolor{lightgray}{00:}13:07                        & 1.32                              \\
\multicolumn{1}{c|}{}                                 & \multicolumn{1}{l|}{{\color[HTML]{009901} \textbf{OOB-decode.}}}           & {\color[HTML]{CB0000} Fail}           & \multicolumn{1}{r|}{\textbf{1}}      & -                                                      & 08:00:43                                               & \textbf{\textgreater6}            \\
\multicolumn{1}{c|}{\multirow{-6}{*}{Gateway}}        & \multicolumn{1}{l|}{{\color[HTML]{009901} \textbf{NPD-processSysex.}}}         & {\color[HTML]{CB0000} Fail}           & \multicolumn{1}{r|}{\textbf{5}}      & -                                                      & 06:30:13                                               & \textbf{\textgreater7.38}         \\ \hline
\multicolumn{1}{c|}{Echo\_Server}                     & \multicolumn{1}{l|}{20-10064\cite{10064descrip}}                               & 90                                    & \multicolumn{1}{r|}{\textbf{95}}     & 03:23:34                                               & 01:38:02                                               & \textbf{2.08}                     \\ \hline
\multicolumn{1}{c|}{Bootstrap(UART)}                  & \multicolumn{1}{l|}{{\color[HTML]{009901} \textbf{NPD-net\_buf\_simple.}}} & {\color[HTML]{CB0000} Fail}           & \multicolumn{1}{r|}{\textbf{1}}      & -                                                      & 05:50:42                                               & \textbf{\textgreater8.23}         \\ \hline
\multicolumn{1}{c|}{Bootstrap(SPI)}                   & \multicolumn{1}{l|}{20-10065\cite{10065descrip}}                               & 10                                    & \multicolumn{1}{r|}{\textbf{15}}     & \textcolor{lightgray}{00:}11:28                        & \textcolor{lightgray}{00:}02:35                        & \textbf{4.44}                     \\ \hline
\multicolumn{1}{c|}{L2cap\_Processor}                 & \multicolumn{1}{l|}{20-12140\cite{12140descrip}}                               & {\color[HTML]{CB0000} Fail}           & \multicolumn{1}{r|}{\textbf{5}}      & -                                                      & \textcolor{lightgray}{00:}53:16                        & \textbf{\textgreater57}           \\ \hline
\multicolumn{1}{c|}{Snmp\_Server}                     & \multicolumn{1}{l|}{20-12141\cite{12141descrip}}                               & \textbf{3}                            & \multicolumn{1}{r|}{2}               & 08:57:22                                               & 06:21:03                                               & 1.41                              \\ \hline
\multicolumn{1}{c|}{}                                 & \multicolumn{1}{l|}{RC-ble\_isr\cite{CCN-Lite-Relay1descrip}}                  & 20                                    & \multicolumn{1}{r|}{\textbf{25}}     & \textcolor{lightgray}{00:}01:12                        & \textcolor{lightgray}{00:}01:46                        & 0.68                              \\
\multicolumn{1}{c|}{}                                 & \multicolumn{1}{l|}{NPD-ccnl\cite{CCN-Lite-Relay2descrip}}                     & {\color[HTML]{CB0000} Fail}           & \multicolumn{1}{r|}{\textbf{2}}      & -                                                      & 18:35:26                                               & \textbf{\textgreater2.69}         \\
\multicolumn{1}{c|}{}                                 & \multicolumn{1}{l|}{NPD-evtimer\cite{CCN-Lite-Relay3descrip}}                  & 4                                     & \multicolumn{1}{r|}{\textbf{5}}      & 09:50:23                                               & 06:11:44                                               & 1.59                              \\
\multicolumn{1}{c|}{\multirow{-4}{*}{CCN-Lite-Relay}} & \multicolumn{1}{l|}{UAF-ccnl\_xx\cite{CCN-Lite-Relay4descrip}}             & 20                                    & \multicolumn{1}{r|}{\textbf{28}}     & 03:49:21                                               & 01:48:32                                               & \textbf{2.11}                     \\ \hline
\multicolumn{1}{c|}{}                                 & \multicolumn{1}{l|}{NPD-net\_pkt\cite{Zephyr_SocketCan2descrip}}               & 6                                     & \multicolumn{1}{r|}{\textbf{8}}      & 16:10:02                                               & 12:37:22                                               & 1.28                              \\
\multicolumn{1}{c|}{\multirow{-2}{*}{SocketCan}}      & \multicolumn{1}{l|}{NPD-pwm\_shell}                                            & 3                                     & \multicolumn{1}{r|}{\textbf{9}}      & 17:42:07                                               & 05:08:16                                               & \textbf{3.45}                     \\ \hline
\multicolumn{1}{c|}{Gattupdate*}                    & \multicolumn{1}{l|}{{\color[HTML]{009901} \textbf{24-22095}}}                   & 40                                    & \multicolumn{1}{r|}{\textbf{43}}     & 07:27:16                                               & \textcolor{lightgray}{00:}15:32                        & \textbf{28.79}                    \\ \hline
\end{tabular}
\end{adjustbox}
\label{tab:eval:bug}
\flushleft
\scriptsize{ 
OOB: Out of Bound Access; ING: Integer Overflow N/UPD: NULL/Uninitialized Pointer Deference; UAF:Use After Free; RC:Race Condition\\ 
*:\texttt{Client-Gattupdate} and \texttt{Server-Gattupdate} samples yield very similar results outcomes, so we present the average result.
}
\end{table}

\subsubsection{Bug Finding Capability Enhancement}
\label{sec:eval:bug}
\sys not only improves coverage but also enhances bug-finding capabilities. We extend the fuzzing test period to over 48 hours to assess the improvement.
As shown in \autoref{tab:eval:bug}, 
\sys triggered crashes more frequently and faster, with some cases being over 10 times quicker than the original method.
For example, \texttt{Bootstrap (SPI)} has an out-of-bound-write vulnerability (CVE-2020-10065) in BLE HCI \texttt{BT\_BUF} buffer. 
Triggering the corresponding crash requires over 77 bytes in a single command to overflow the \texttt{BT\_BUF} buffer. If a command is received before HCI initialization or during the last command processing stage, it aborts the current command and responds with an error(\texttt{HCI\_ERROR\_CMD\_DISALLOWED}).
However, the RR or fuzz delivery strategy used by \mulitfuzz and \fuzzware delivers only one byte of input at uncertain time, often causing the HCI controller to abort the retrieved command.
This reduces the chance of accumulating a long input that could cause a buffer overflow, making it difficult for the fuzzer to detect this vulnerability. 
In contrast, \sys delivers all inputs based on firmware requirements (\ie~at checking HCI command availability point), allowing longer command retrieval and increasing the likelihood of triggering the vulnerability (7min to trigger with \sys vs. 7h with RR).

\paragraph{Analysis of newly Discovered Bugs.}
\sys uniquely detected ten bugs (two for \semu, two for \fuzzware, and six for \mulitfuzz) including five 0-day bugs missed by RR and Fuzz delivery methods even in prior work's extensively fuzzed samples.

In the \texttt{Gateway}, two new bugs were found in the I2C input route processing functions. 
The first is a buffer overflow during decoding of longer messages. The second is a NULL pointer dereference, occurring when a command lacks a parameter field, causing the input buffer allocation of parameter field to return NULL and leading to a system crash due to uninitialized NULL pointer dereference. \sys can uniquely identify these bugs by enabling input processing for I2C routes through solving P4, as mentioned later, while RR and Fuzz delivery failed. The unique detection of known bugs in \texttt{CCN-Lite-Relay}, \texttt{Heat\_Press}, and \texttt{PLC} also shares the same reasons.

The \texttt{Client/Server-Gattupdate} samples vulnerability is in BLE Cordio implementation of Arm Mbed OS 6.17.0, occur when an invalid packet type is received, leading to buffer overflow due to data accumulation 
in a constrained while loop.
The \sys-enable fuzzer effectively delivers inputs in a stable pattern, 
allowing long inputs to fill in the buffer and trigger overflow. 
In contrast, the RR and Fuzz delivery method feeds inputs separately at incorrect timings, 
causing frequent buffer resets due to error handling, thus reducing the likelihood of triggering the overflow.

The \texttt{BLE-HCI} vulnerability in the BLE subsystem implementation of Zephyr 4.1.0 arises from an uninitialized connection address variable in the memory pool during BLE initialization, leading to a device crash when dereferenced later in the command response process (\texttt{tx\_demux}).
To trigger this crash, BLE must complete initialization and retrieve the command first.
\sys uniquely detected this bug by enabling more times and longer input retrieval, similar to the CVE-2020-10065.

The \texttt{BLE-Bootstrap(UART)} vulnerability occurs when the allocated space for input data exceeds the remaining space, causing the allocation to fail and return a NULL pointer buffer, leading to a NULL pointer dereference. Under RR and Fuzz, the delivery interval is long, making it difficult to input to accumulated in heap. However, \sys can deliver longer input at once at delivery point, allowing this bug to be detected.

\begin{table}[t]
\caption{Issues with \aid under frequency 1 (F=1) and frequency 10 (F=10).(-: The execution is stuck.)}
\centering
\begin{adjustbox}{max width=\columnwidth}
\begin{tabular}{l|c|cc|cc}
\hline
\multicolumn{1}{c|}{\multirow{2}{*}{\textbf{Firmware}}} & \multicolumn{1}{c|}{\textbf{Unsatisfied Input Route}}        & \multicolumn{2}{c|}{\textbf{F = 1}}           & \multicolumn{2}{c}{\textbf{F =10}} \\
\multicolumn{1}{c|}{}                                   & \multicolumn{1}{c|}{\textbf{Perip.:\#:[Lower:Upper],...}} & \textbf{P1} & \multicolumn{1}{c|}{\textbf{P2}} & \textbf{P3}      & \textbf{P4}      \\ \hline
Console                                                 & UART:1:[1:64]& No          & No                               & Yes              & No               \\
Steering\_Control                                       & UART:2:[1:128]& Yes         & No                               & Yes              & No               \\
Gateway                                                 & UART:1:[1:64],I2C:2:[1:128]& No          & No                               & Yes              & Yes              \\
Heat\_Press                                             & UART:6:[8:64]& -           & Yes                              & Yes              & No               \\
PLC                                                     & UART:1:[8:64]& -           & Yes                              & Yes              & No               \\
Soldering\_Iron                                         & I2C:1:[1:128]& No          & No                               & Yes              & No               \\
GPS\_Tracker                                            & UART:1:[1:256],UART:4:[1:256]& No          & No                               & Yes              & Yes              \\
LiteOS\_IoT                                             & UART:1:[1:100],UART:5:[1:100]& No          & No                               & Yes              & Yes              \\
3Dprinter                                               & UART:1:[1:64],USB:1& No          & No                               & Yes              & No               \\
SocketCan                                               & CAN:1:[1:64]& No          & No                               & Yes              & No               \\
$\mu$Tasker\_USB                                        & USB:1:[1:512]& No          & No                               & Yes              & No               \\
Bootstrap(UART)                                         & UART:1:[1:5]& Yes         & No                               & Yes              & No               \\
Bootstrap(SPI)                                          & SPI:1:[1:5]& Yes         & No                               & Yes              & No               \\
Echo\_Server                                            & SPI:1:[1:132]& No          & No                               & Yes              & No               \\
L2cap\_Processor                                        & RADIO:1:[1:128]& No          & No                               & Yes              & No               \\
Snmp\_Server                                            & RADIO:1:[1:128]& No          & No                               & Yes              & No               \\
CCN-Lite-Relay                                          & UART:1:[1:128],Radio:1:[1:168]& No          & No                               & Yes              & Yes              \\
Gnrc\_networking                                        & UART:2:[1:128]& No          & No                               & Yes              & No               \\
Client-Gattupdate                                       & SPI:2:[1:256]& No          & No                               & Yes              & No               \\
Server-Gattupdate                                       & SPI:2:[1:256]& No          & No                               & Yes              & No               \\
BLE-HCI                                                 & UART:1:[1:7]& Yes         & No                               & Yes              & No               \\ \hline
\end{tabular}
\end{adjustbox}
\label{tab:aidfuzzer}
\end{table}

\subsection{Comparison to Interrupt-driven Firmware Fuzzers (RQ3)}
\label{sec:eval:aid}

\begin{table*}[t]
\caption{Ablation study for the effect of S1-S3, using \fuzzware in RR mode as
the baseline. $\uparrow$ indicates changes in median coverage and average crash count
compared to the previous configuration. Changes below 0.1\% are not displayed,
and significant changes are marked in bold (based on a Mann-Whitney U test with
a 0.05 significance threshold).}
\centering
\begin{adjustbox}{max width=\textwidth}
\begin{tabular}{l|cc|rlcrl|rlcrl|rlcrl}
\hline
\multicolumn{1}{c|}{}                                    & \multicolumn{2}{c|}{\textbf{RR}}                    & \multicolumn{5}{c|}{\textbf{S1}}                                                                                                                                                           & \multicolumn{5}{c|}{\textbf{+S2}}                                                                                                                                                          & \multicolumn{5}{c}{\textbf{+S3}}                                                                                                                                                 \\
\multicolumn{1}{c|}{}                                    & \multicolumn{1}{c|}{\textbf{Cov.}} & \textbf{Crash} & \multicolumn{3}{c|}{\textbf{Coverage}}                                                                                & \multicolumn{2}{c|}{\textbf{Crash}}                                & \multicolumn{3}{c|}{\textbf{Coverage}}                                                                                & \multicolumn{2}{c|}{\textbf{Crash}}                                & \multicolumn{3}{c|}{\textbf{Coverage}}                                                                                & \multicolumn{2}{c}{\textbf{Crash}}                       \\
\multicolumn{1}{c|}{\multirow{-3}{*}{\textbf{Firmware}}} & \multicolumn{1}{c|}{\textbf{Med.}} & \textbf{Count} & \textbf{Med.}           & \multicolumn{1}{c}{\textbf{$\uparrow$}} & \multicolumn{1}{c|}{\textit{p-value}}             & \textbf{Count}          & \multicolumn{1}{c|}{\textbf{$\uparrow$}} & \textbf{Med.}           & \multicolumn{1}{c}{\textbf{$\uparrow$}} & \multicolumn{1}{c|}{\textit{p-value}}             & \textbf{Count}          & \multicolumn{1}{c|}{\textbf{$\uparrow$}} & \textbf{Med.}           & \multicolumn{1}{c}{\textbf{$\uparrow$}} & \multicolumn{1}{c|}{\textit{p-value}}             & \textbf{Count} & \multicolumn{1}{c}{\textbf{$\uparrow$}} \\ \hline
Gateway                                                  & \multicolumn{1}{c|}{2362}          & 0              & 2,512                   & +6.4\%                                  & \multicolumn{1}{c|}{{\color[HTML]{9B9B9B} 0.056}} & 0                       &                                          & 2,558                   & +1.8\%                                  & \multicolumn{1}{c|}{{\color[HTML]{9B9B9B} 0.548}} & 0                       &                                          & 2,756                   & +7.7\%                                  & \multicolumn{1}{c|}{{\color[HTML]{9B9B9B} 0.095}} & 0              &                                         \\
Heat\_Press                                              & \multicolumn{1}{c|}{551}           & 247.2          & \multicolumn{1}{r}{563} & +2.2\%                                   & \multicolumn{1}{c|}{0.010}                        & \multicolumn{1}{r}{235.6} & -4.7\%& \multicolumn{1}{r}{565} & +0.4\%                                   & \multicolumn{1}{c|}{\color[HTML]{9B9B9B} 0.666}                        & \multicolumn{1}{r}{259} & +9.9\%& \multicolumn{1}{r}{570} & +0.9\%                                   & \multicolumn{1}{c|}{\color[HTML]{9B9B9B} 0.916}                        & 353            & +36.3\%                                  \\
CCN-Lite-Relay                                           & \multicolumn{1}{c|}{491}           & 0              & 1,054                   & +114.7\%                                & \multicolumn{1}{c|}{0.011}                        & 0                       &                                          & 1,054                   &                                         & \multicolumn{1}{c|}{{\color[HTML]{9B9B9B} 0.796}} & 0                       &                                          & 1,054                   &                                         & \multicolumn{1}{c|}{{\color[HTML]{9B9B9B} 1.000}} & 0              &                                         \\
Gnrc\_Networking                                         & \multicolumn{1}{c|}{421}           & 0              & 666                     & +58.2\%                                 & \multicolumn{1}{c|}{0.014}                        & 0                       &                                          & 666                     &                                         & \multicolumn{1}{c|}{{\color[HTML]{9B9B9B} 0.821}} & 0                       &                                          & 668                     & +0.3\%                                  & \multicolumn{1}{c|}{{\color[HTML]{9B9B9B} 0.564}} & 0              &                                         \\
3Dprinter                                                & \multicolumn{1}{c|}{786}           & 0              & 902                     & +14.8\%                                 & \multicolumn{1}{c|}{{\color[HTML]{9B9B9B} 0.056}} & 0                       &                                          & 901                     &                                         & \multicolumn{1}{c|}{{\color[HTML]{9B9B9B} 0.916}} & 0                       &                                          & 931                     & +3.3\%                                  & \multicolumn{1}{c|}{{\color[HTML]{9B9B9B} 0.140}} & 0              &                                         \\
GPSTracker                                               & \multicolumn{1}{c|}{661}           & 0              & 948                     & +43.4\%                                 & \multicolumn{1}{c|}{0.032}                        & 0                       &                                          & 952                     & +0.4\%                                  & \multicolumn{1}{c|}{{\color[HTML]{9B9B9B} 1.000}} & 0                       &                                          & 1,011                   & +6.2\%                                  & \multicolumn{1}{c|}{{\color[HTML]{373C43} 0.012}} & 0              &                                         \\
LiteOS\_IoT                                              & \multicolumn{1}{c|}{738}           & 0              & 739                     & +0.1\%                                  & \multicolumn{1}{c|}{{\color[HTML]{9B9B9B} 0.083}} & 0                       &                                          & 1,079                   & +46.0\%                                 & \multicolumn{1}{c|}{{\color[HTML]{9B9B9B} 0.408}} & 0                       &                                          & 1,333                   & +23.5\%                                 & \multicolumn{1}{c|}{{\color[HTML]{9B9B9B} 0.292}} & 0              &                                         \\ \hline
\end{tabular}
\end{adjustbox}
\label{tab:eval:abla}

\end{table*}

A recent work, \aid~\cite{wangaidfuzzer}, introduces an interrupt-driven
test-case delivery mechanism, aiming to address a problem similar to that of
\sys. The main observation of \aid is that firmware often enters waiting states,
such as when encountering sleep instructions like \texttt{WFI}, entering an
infinite loop, or reading a global variable modifiable in an ISR. \aid delivers
input when a waiting state is detected. While \sys is designed as a drop-in
replacement for the input delivery mechanism of existing fuzzers, we found it
challenging to integrate \sys with \aid's underlying emulator due to the tight
coupling of \aid's interrupt mechanism with its emulator.

To compare \sys with \aid, we conducted two experiments. First, we tested 27 
samples using \aid and compared the results with \sys using \mulitfuzz. 

After 24
hours, \aid produced output for 16 samples, while execution failed for others
due to unsupported memory map alignment or initial seed crashes. For the 16
successful samples, \sys with \mulitfuzz consistently achieved significantly
higher code coverage, ranging from 5\% to over 1,500\%, as detailed in
Appendix~\ref{app:cov}.

In our second experiment, we re-implemented one of \aid's waiting-state
detections (reading a global variable modifiable in an ISR) and integrated it
into \mulitfuzz. Additionally, \aid employs an opportunistic strategy with a
configurable triggering frequency, randomly skipping some waiting-state
encounters. Our configuration covers both the minimum and maximum frequencies
(1/10). The results are summarized in~\autoref{tab:aidfuzzer}.

\paragraph{Problems under Minimum Frequency.}
When the ISR trigger configuration time is set to one, \aid delivers input
whenever a waiting state is encountered, without considering the buffer's
maximum capacity, which causes P1. For example, \texttt{Steering\_Control} reads
input until it encounters a comma, line break, or an empty buffer.  Before each
read, it checks global variables. Consequently, \aid continues to deliver input
before buffer reading, regardless of whether the input length exceeds the buffer
capacity. Additionally, as mentioned in F2, some firmware has buffer cleaning
behavior ($\widetilde{R_{B}}$), and input should not be delivered at
$\widetilde{R_{B}}$, as it would be wasted. However, if the frequency is one,
\aid can avoid this issue. For instance, in \autoref{lst:heatpress}, \aid
continuously feeds new data into the buffer at the \texttt{Head} pointer
reading, causing the firmware to become stuck in the data retrieval loop and
waste input (P2) (\autoref{line:check2}).

\paragraph{Problems under Maximum Frequency.}
For \aid, if the configuration frequency exceeds 1, the firmware enters the
waiting state at least twice, but only one time interrupt is triggered
(\ie~typically, one byte is delivered per interrupt). This results in failed
availability checks in P3 and P4. Taking \texttt{Heat\_Press} as an example
in~\autoref{lst:heatpress}, since the \texttt{Head} pointer is modified in the
ISR, \aid interprets \texttt{Head} pointer readings in functions like
\texttt{available} as waiting states. If the waiting state is set to ten, only
one byte can be retrieved from the ISR buffer after ten availability checks
at~\autoref{line:check2}, preventing the length check from passing.
Additionally, not all availability checks for each \CRProute are unconditional
repeated; some routes are checked only under specific conditions, such as in
\texttt{Gateway}, \texttt{GPS\_Tracker}, and \texttt{CCN-Lite-Relay}. A
high-frequency configuration (with less interrupt triggering) can lead to P4
problems.

In summary, we found that the coarse-grained input delivery reasoning in \aid
mitigates the issue with random input delivery to some extent, but it cannot
effectively address all the identified problems in this paper.


\subsection{Ablation Study (RQ4)}
\label{sec:eval:ablation}

To quantify the marginal contribution of each component, including delivery
point timing inference (S1), length-range inference (S2), and multi-route-aware
delivery coordination (S3), we conduct ablation studies using fuzzing metrics
(\ie~basic block coverage and crash counts). These studies are performed on
representative samples involving multiple input routes (\texttt{3DPrinter},
\texttt{GPSTrackers}, \texttt{CCN-Lite-Relay}, \texttt{Gnrc\_Networking},
\texttt{Gateway}, \texttt{Heat\_Press}, and \texttt{LiteOS\_IoT}; see details in
\autoref{tab:detailsample} in the Appendix) to demonstrate the impact of each
component. To isolate the effects of each component, we start with a baseline
configuration using Fuzzware in RR delivery mode. We then construct three
configurations for the ablation study, each incrementally adding one component
of \sys. The first configuration includes only S1, the second adds S2 to S1, and
the third includes all three components (S1, S2, and S3). The results are shown
in~\autoref{tab:eval:abla}.

\paragraph{+S1}: We utilize delivery point information extracted by S1 for
fuzzing, but deliver only one byte at each point via interrupt. Enabling S1
improves coverage by 34.3\% on average by avoiding P4. In contrast, interrupt
delivery via round-robin in \fuzzware is less effective because most interrupt
triggering does not deliver any byte (\eg~\texttt{Gateway},
\texttt{CCN-Lite-Relay}, and \texttt{GPSTracker}, detailed in
Appendix~\ref{app:detailinfo}). For samples with length checks
(\eg~\texttt{LiteOS\_IoT} with a 20B limit and \texttt{Heat\_Press} with 8B), S1
alone requires more time to reach the limit, resulting in less coverage
improvement and slightly fewer crashes than the round-robin mode for
\texttt{Heat\_Press}.

\paragraph{+S2}: Building on S1, S2 ensures that the delivery length meets the
required ranges, effectively handling samples such as \texttt{LiteOS\_IoT} and
\texttt{Heat\_Press} with length checks. This increases coverage for
\texttt{LiteOS\_IoT} by 46\% and crash counts for \texttt{Heat\_Press} by 10\%
compared to S1 alone. For other samples, length checks are not applicable;
therefore, S2 does not provide improvement. Without a delivery schedule (S3), S2
alone may introduce negative effects because the distribution of input routes
per fuzzing round becomes unpredictable, leading to uneven input distribution
(\eg~\texttt{LiteOS\_IoT} and \texttt{Heat\_Press}).

\paragraph{+S3}: Enabling S3 offers two benefits: (1) maximizing coverage by
testing all input routes (achieving a 6\% average increase on top of S1+S2), and
(2) mitigating the negative effects of S2 by balancing input distribution. For
example, a vulnerability in \texttt{Heat\_Press} was found in one of six routes.
The input scheduling algorithm~\autoref{alg:InputSliceAlg} reserves a minimum
input length for each route, ensuring stable delivery across all routes and
increasing crash numbers by 36\% compared to S1+S2.

%% file: tex/7-discuss.tex
\section{Limitations and Discussion}

\label{sec:discuss:inter}

\paragraph{Non-IO Handling.}
Our design focuses on how firmware handles inputs from real peripherals, meaning
that fuzzing inputs should only be consumed by external data register reads.
Internal peripheral registers, such as status registers, should be managed by
real hardware or emulation. \sys relies on existing firmware fuzzers in handling
internal peripheral registers, and thus inherits their limitation to use fuzzing
inputs for both data register reads and some internal hardware register reads.
This can lead to minor inaccuracies in sub-input length calculation in Step~3,
causing P1--P4 problems to occur.


\paragraph{Orthogonality with Existing Front-end Optimizations.}
In the fuzzer front-end, the proposed input delivery is not entirely orthogonal
from existing front-end optimizations such as input-to-state (I2S) and length
extension. They all influence whether mutated inputs reach and exercise the
processing logic, and their interaction can be complex. For instance, an I2S
mutation might adjust byte positions for long-string comparisons; if delivery
timing and length change, the effective byte positions may shift, reducing
effectiveness. It is our future work to more thoroughly study the interactions
when integrating delivery information into existing fuzzing optimizations.

\paragraph{DMA Support.}
High-throughput peripherals like USB and Ethernet commonly use DMA to allow data
transfers between RAM and peripherals without processor involvement. \sys as a
plugin for \mulitfuzz, \fuzzware and \semu, does not directly support emulating
DMA peripherals. To support DMA, we need to identify the delivery point for DMA
transactions. Inspired by GDMA~\cite{scharnowski2025gdma}, we found that \sys's
analysis can be ported to achieve DMA delivery point identification, with some
chip-specific knowledge. For example, the six common DMA configurations
summarized in GDMA can also be mapped to two notification modes similar to
polling and interrupt modes. 

In the polling mode, after a DMA transaction, specific fields in MMIO registers
(MMIO-based DMA Configuration) or DMA descriptors (RAM-based DMA Configuration)
are updated. Firmware checks this field to determine input availability, similar
to polling mode checks. In the interrupt mode, DMA can be set to trigger an
interrupt upon transaction completion, updating global variables. Firmware then
checks these variables to confirm transaction completion, similar to interrupt
mode. Therefore, we can reuse \sys to identify global variable checking points
as the delivery points for DMA data. However, determining delivery length,
usually indicated by specific fields in MMIO registers or descriptors, requires
precise DMA modeling like GDMA, which our length inference method cannot
achieve.


\paragraph{Generalizability.}
While the idea of delivery information extraction is general to firmware, its
implementation is specific to the underlying fuzzer's front-end or back-end.


%% file: tex/8-related_work.tex
\section{Related Work}

\paragraph{Firmware Dynamic Analysis.}
Dynamic analysis techniques, particularly fuzzing, are effective in identifying bugs in various software~\cite{zhu2022fuzzing}. However, applying these techniques to MCU firmware is challenging due to the reliance on resource-constrained hardware and the lack of source code. Some researchers have tried integrating these techniques with original hardware~\cite{talebi2018charm,zaddach2014avatar,muench2018avatar,li2022muafl,liu2024co3,mera2024shift}. 
Testing with physical hardware is challenging for scaling and is often ineffective. Consequently, recent efforts focus on developing effective re-hosting environments for firmware fuzzing~\cite{clements2020halucinator, gustafson2019toward,zhou2021automatic,tobias2022fuzzware,fengp2020p2im,spensky2021conware,chong2024afriend}.
For instance, \fuzzware~\cite{tobias2022fuzzware} and \uemu~\cite{zhou2021automatic} used symbolic execution to derive peripheral MMIO models from firmware behaviors for emulating peripheral reads.
\semu~\cite{zhou2022your} and \textit{Perry}~\cite{chong2024afriend} created peripheral models by extracting hardware logic from public manuals or peripheral driver code, to increase the emulation fidelity.


On the other hand, recent researchers have begun adapting fuzzing techniques to incorporate these features.
\emberio~\cite{farrelly2023ember}  remaps edge-coverage feedback to eliminate invalid new edges caused by random interrupts, while \textit{SplITS}~\cite{farrelly2023splits} addresses multi-byte magic strings in firmware to uncover new code coverage faster.
\hoedur~\cite{scharnowski2023hoedur} and \mulitfuzz~\cite{chesser2024multifuzz} adapt general fuzzing techniques, including input generation, mutation, and feedback, to account for the multi-stream nature of firmware inputs from various peripherals.
In comparison, we identify asynchronous input handling as a new feature impacting firmware fuzzing and propose \sys to improve input delivery. 
In addition, since \sys focuses solely on enhancing input delivery, it can be used alongside other state-of-the-art firmware fuzzing optimizations and emulations.
\aim~\cite{feng2023aim} predicts interrupt-firing timing using persistent symbolic execution, but it is incompatible with fuzzing and incurs significant overhead from symbolic execution.
\aid~\cite{wangaidfuzzer} suggests a waiting state-based interrupt firing solution. However, delivering at every waiting state may stuff the firmware, whereas delivering too infrequently may starve it.
Moreover, \aid does not attempt to determine the proper amount of data to provide, leaving the delivery quantity undefined.



\paragraph{Firmware Static Analysis.}
Static analysis typically used in firmware security analysis to identify specific vulnerabilities.
For example, 
Firmalice~\cite{shoshitaishvili2015firmalice} and PASAN~\cite{kim2021pasan} target authentication bypass and race condition and peripheral access. 
SaTC~\cite{chen2021sharing} employs static data-flow analysis to detect taint-style vulnerabilities by identifying user input through shared keywords.
Additionally, static analysis aids firmware fuzzing; for instance, SFuzz~\cite{chen2022sfuzz} uses forward slicing to prune paths that are independent of external inputs, addressing firmware emulation challenges. Our tool, \sys, aligns with this approach by utilizing static data and control flow analysis to identify optimal input delivery timing and quantity for fuzzing.



%% file: tex/9-con.tex
\section{Conclusion}
In this work, we found that the delivery method also impacts effectiveness and efficiency due to asynchronous interactions between peripheral input arrival and firmware input handling, a factor often overlooked. 
To identify the optimal delivery time and quantity,
We developed, \sys, which automatically extracts delivery information---such as delivering points and expected input volume range---by identifying and analysis the semantic of key input handling operations in firmware programming model (check-retrieval-processing, CRP) through static and dynamic analysis. 
Integrated with SOTA firmware fuzzer, we show that \sys significantly improves fuzzing by increasing coverage and capability of bug detection compared to ad-hoc delivery method such as periodic, fuzz and \MSP pattern. 
Our findings highlight the importance of firmware-aware input delivery mechanisms in firmware fuzzing and open new area for firmware fuzzing improvement.

%% file: tex/11-app.tex

\section{Feature of Unit-test Samples}


\label{app:feature}
\begin{table}[H]

\caption{Feature of unit-test samples (-:No official demo)}
\centering
\begin{adjustbox}{width=\columnwidth}
\begin{tabular}{lccccc}
\hline
\multicolumn{6}{c}{\textbf{RTOS Driver Demo}}                                                                                                           \\ \hline \hline 
\multicolumn{1}{l|}{}              & \textbf{RIoT}         & \textbf{Nuttx}          & \textbf{Zephyr}      & \textbf{MbedOS}  & \textbf{FreeRTOS} \\ \cline{2-6} 
\multicolumn{1}{l|}{\textbf{GPIO}} & F3            & F2, F3              & F3               & F3           & F3            \\
\multicolumn{1}{l|}{\textbf{UART}} & F1, F2, F3            & F1, F3              & F1, F3               & F1, F3           & F1, F3            \\
\multicolumn{1}{l|}{\textbf{I2C}}  & F1, F3            & F1, F3              & F1, F3               & F1, F3           & F1, F2, F3            \\
\multicolumn{1}{l|}{\textbf{SPI}}  & -            & -              & F1, F3               & F1, F3           & F1, F3            \\
\multicolumn{1}{l|}{\textbf{ADC}}  & -            & F2, F3              & F1, F3               & F1, F3           & F1, F3            \\
\multicolumn{1}{l|}{\textbf{ETH}}  & F2, F3            & F2, F3              & F3               & F3           & F3            \\ \hline
\multicolumn{6}{c}{\textbf{Bare-Metal SDK Demo}}                                                                                                              \\ \hline \hline 
\multicolumn{1}{l|}{}              & \multicolumn{2}{c}{\textbf{STM32F103/F429/L152 HAL}} & \textbf{NXP K64/K66} & \multicolumn{2}{c}{\textbf{Microchip SAM3}}   \\
\multicolumn{1}{l|}{}              & \textbf{ARDUINO}      & \textbf{STM32Cude}      & \textbf{HAL SDK}     & \textbf{ARDUINO} & \textbf{HAL SDK}      \\ \cline{2-6} 
\multicolumn{1}{l|}{\textbf{GPIO}} & F2, F3            & F3              & F3               & F2, F3           & F3            \\
\multicolumn{1}{l|}{\textbf{UART}} & F1, F2, F3            & F1, F3              & F1, F3               & F1, F3           & F1, F3            \\
\multicolumn{1}{l|}{\textbf{I2C}}  & F1, F2, F3            & F1, F3              & F1, F3               & F1, F2, F3           & F1, F2 ,F3            \\
\multicolumn{1}{l|}{\textbf{SPI}}  & F1, F3            & F1, F3              & F1, F3               & F1, F3           & F1, F2, F3            \\
\multicolumn{1}{l|}{\textbf{ADC}}  & F2, F3            & F1, F3              & F3               & F1, F3           & F1, F3            \\
\multicolumn{1}{l|}{\textbf{ETH}}  & F2, F3            & F2, F3              & F3               & F2, F3           & F3            \\ \hline
\end{tabular}
\end{adjustbox}
\label{tab:unitfeature}

\end{table}




\section{Detail Configuration of Demo in~\autoref{sec:problems}}
\label{app:config}

Configuration-1 used a round-robin delivery method, delivering input every 1,000 basic blocks executed (the default for \fuzzware).
Configuration-2 employed a fuzz delivery method, with intervals ranging from 1 to 16,000 basic blocks executed, 
increasing in steps of 250 times powers of 2 from zero to six. 
One interval is chosen based on the fuzzing input data modulo eight. 

For these three groups, the delivery length varied: 1B (the one-time DR read length for the \texttt{UART\_IRQhandler} function, 
the default for \fuzzware), a random size between 1 and 1,000B (to mimic unconstrained fuzzer-generated input lengths), and 
a manually restricted length between 8 and 127 (to satisfy minimum length checks).

Configuration-3 uses fixed input delivery points at the start of the main \texttt{Loop} function (\autoref{line:loop} in~\autoref{lst:heatpress}) with a random size between 1 and 1,000B (to mimic unconstrained fuzzer-generated input lengths). 

Configuration-4 simulates the ideal situation that delivers inputs at the start of the \texttt{Poll} 
function with sizes randomly selected within the specified range (8 to 127).

\section{Delivery Information of Unit-test Samples}
\label{app:dpinfo}

\begin{table}[h]
\caption{Delivery information of unit-test samples}
\centering
\begin{adjustbox}{width=\columnwidth}
\begin{tabular}{lrrrr}
\hline
\multicolumn{5}{c}{\textbf{RTOS Driver Demos (CRP Route(Perip.:\#:[Lower:Upper]))}}                                                                                                                            \\ \hline
\multicolumn{1}{c}{}               & \multicolumn{1}{c}{\textbf{F103/RIOT}} & \multicolumn{1}{c}{\textbf{F103/NUTTX}} & \multicolumn{1}{c}{\textbf{K64/RIOT}}     & \multicolumn{1}{c}{\textbf{SAM3/RIOT}}    \\ \hline
\multicolumn{1}{l|}{\textbf{GPIO}} & GPIO:1& GPIO:1& GPIO:1& GPIO:1\\
\multicolumn{1}{l|}{\textbf{UART}} & UART:1:[1:64]& UART:1:[1:12]& UART:1:[1:64]& UART:1:[1:64]\\
\multicolumn{1}{l|}{\textbf{I2C}}  & I2C:1:[1:4]& I2C:1:[1:32]& -                                         & -                                         \\
\multicolumn{1}{l|}{\textbf{ADC}}  & -                                      & ADC:1:[1:4]& ADC:1:[1:4]& -                                         \\ \hline
\multicolumn{5}{c}{\textbf{Bare-Metal  Demos (CRP Route(Perip.:\#:[Lower:Upper]))}}                                                                                                                            \\ \hline
                                   & \multicolumn{1}{c}{\textbf{K64/HAL}}   & \multicolumn{1}{c}{\textbf{K66/HAL}}    & \multicolumn{1}{c}{\textbf{F103/Arduino}} & \multicolumn{1}{c}{\textbf{SAM3/Arduino}} \\ \hline
\multicolumn{1}{l|}{\textbf{GPIO}} & GPIO:1& GPIO:1& GPIO:1& GPIO:1\\
\multicolumn{1}{l|}{\textbf{UART}} & UART:1:[1:64]& UART:1:[1:64]& UART:1:[1:128]& UART:1:[1:128]\\
\multicolumn{1}{l|}{\textbf{I2C}}  & I2C:1:[1:100]& I2C:1:[1:100]& I2C:1:[1:100]& I2C:1:[1:100]\\
\multicolumn{1}{l|}{\textbf{ADC}}  & ADC:1:[1:4]& ADC:1:[1:4]& ADC:1:[1:4]& ADC:1:[1:4]\\ \hline                 
\end{tabular}
\end{adjustbox}
\label{tab:detailunittest}
\end{table}

\section{Responsible Disclosure}
\label{app:bugstatus}

As of March 31, 2026, the bugs found in \texttt{Client/Sever-Gattupdate} and
\texttt{BLE-HCL} have been fixed. The bug in \texttt{Bootstrap(UART)} was
acknowledged by the vendor, and a fix is under discussion. Two bugs in
\texttt{Gateway}, which are only locally exploitable, have not received any
response from the vendor.


\begin{table}[h]
\caption{Detail of newly discovered Bugs by \sys}
\centering
\begin{adjustbox}{width=\columnwidth}
\begin{tabular}{l|l|l|l}
\hline
\multicolumn{1}{c|}{\textbf{Firmware}} & \multicolumn{1}{c|}{\textbf{Bug Type}} & \multicolumn{1}{c|}{\textbf{Vulnerable Func.}} & \multicolumn{1}{c}{\textbf{Status}} \\ \hline \hline
BLE-HCL                                & NULL Pointer Deference                 & ull\_conn\_tx\_lll\_enqueue                    & Fixed                               \\
Gateway                                & Out-of-bound-write                     & decodeByteStream                               & Reported                            \\
Gateway                                & NULL Pointer Deference                 & processSysexMessage                            & Reported                            \\
Bootstrap(UART)                        & NULL Pointer Deference                 & net\_buf\_simple\_tailroom                     & Acknowledged                        \\
Client/Sever-Gattupdate                & Out-of-bound-write                     & hciTrSerialRxIncoming                          & Fixed                               \\ \hline
\end{tabular}         
\end{adjustbox}
\label{tab:detailbug}
\end{table}

\section{Details of Real-world Firmware Samples}

\label{app:detailinfo}
\begin{table}[h!]
\caption{Details and input routes of 25 real-world firmware samples (Underline input
routes in polling mode, others in interrupt mode. Bold input routes that always occur in main loop; other routes occur only in specific conditions.)}
\centering
\begin{adjustbox}{width=\columnwidth}
\begin{tabular}{l|l|c|c|c}
\hline
\textbf{Firmware}          & \multicolumn{1}{c}{\textbf{MCU}} & \textbf{OS/Sys lib.} & \textbf{Total} & \textbf{CRP Route(Peripheral.:\#:[Lower:Upper],...))}                              \\ \hline \hline
Console                    & NXP K64F                         & NXP HAL              & 2,251          & \textbf{UART:1:[1:64]}                                                        \\
Steering\_Control          & SAM3X                            & Arduino              & 1,835          & \textbf{UART:2:[1:128]}                                                       \\
Gateway                    & STM32F103                        & Arduino              & 4,921          & \textbf{UART:1:[1:64]},I2C:2:[1:128],\underline{GPIO:4},\underline{ADC:1}     \\
Heat\_Press                & SAM3X                            & Arduino              & 1,837          & \textbf{UART:6:[8:64]}                                                        \\
PLC                        & STM32F429                        & Arduino              & 2,304          & \textbf{UART:1:[8:64]}                                                        \\
Soldering\_Iron            & STM32F103                        & FreeRTOS             & 3,657          & \textbf{I2C:1:[1:128]},\underline{\textbf{GPIO:1}},\underline{\textbf{ADC:1}} \\
GPS\_Tracker               & SAM3X                            & Arduino              & 4,194          & \textbf{UART:1:[1:256]},UART:4:[1:256]                                        \\
LiteOS\_IoT                & STM32L431                        & LiteOS               & 2,423          & \textbf{UART:1:[20:100]},UART:5:[20:100]                                        \\
3Dprinter                  & STM32F103                        & STM32 HAL            & 8,045          & \textbf{UART:1:[1:64]},\textbf{USB:1},\underline{GPIO:2}                      \\
SocketCan                  & STM32L432                        & Zephyr               & 5,943          & \textbf{CAN:1:[1:64]}                                                         \\
$\mu$Tasker\_USB           & STM32F429                        & $\mu$Tasker          & 3,491          & \underline{UART:1:[1:516]},USB:1:[1:512],\underline{\textbf{GPIO:1}}          \\
Bootstrap(UART)            & nRF52840                         & Zephyr               & 4,972          & \textbf{UART:1:[1:5]},\underline{GPIO:1}                                      \\
Bootstrap(SPI)             & nRF52840                         & Zephyr               & 4,949          & \textbf{SPI:1:[1:5]},\underline{GPIO:1}                                       \\
Echo\_Server               & SAM4E                            & Zephyr               & 7,007          & \textbf{SPI:1:[1:132]}                                                        \\
L2cap\_Processor           & TICC2538                         & Contiki-NG           & 4,002          & \textbf{RADIO:1:[1:128]}                                                      \\
Snmp\_Server               & TICC2538                         & Contiki-NG           & 3,080          & \textbf{RADIO:1:[1:128]}                                                      \\
CCN-Lite-Relay             & nRF52832                         & Nordic HAL           & 12,675         & \textbf{UART:1:[1:128]},   Radio:1:[1:168]                           \\
Gnrc\_networking           & STM32F303                        & STM32 HAL            & 6,448          & \textbf{UART:2:[1:128]}                                                       \\
Filesystem                 & STM32F303                        & STM32 HAL            & 2,429          & \textbf{UART:1:[1:128]},UART:4:[1:128]                                        \\

6Lowpan\_Receiver           & SAM R21                         & Contiki               & 6,988          & UART:1, \textbf{\underline{Radio:1}}, \underline{I2C:1}                                                       \\
6Lowpan\_Transmitter           & SAM R21                         & Contiki               & 6,988          & UART:1, \textbf{\underline{Radio:1}}, \underline{I2C:1}                                                        \\ 
P2IM\_Drone           & STM32F103                         & Bare-Metal               & 2,754          & UART:1:[1:2], \textbf{\underline{I2C:4}}                                                       \\ \hline
\textbf{Client-Gattupdate} & nRF52840                         & MBedOS               & 13,888         & \textbf{SPI:2:[1:256]},\underline{\textbf{GPIO:1}}                            \\
\textbf{Server-Gattupdate} & nRF52840                         & MBedOS               & 13,826         & \textbf{SPI:2:[1:256]},\underline{\textbf{GPIO:1}}                            \\
\textbf{BLE-HCI}           & nRF52840                         & Zephyr               & 7,470          & \textbf{UART:1:[1:7]}                                                         \\ 
\hline
\end{tabular}
\end{adjustbox}
\label{tab:detailsample}
\end{table}

\onecolumn

\section{Detail of Fuzzing Coverage Comparison Results}
\label{app:cov}

\begin{figure*}[h]
    \raggedright   
   
    \begin{subfigure}[b]{\textwidth}
        \raggedright
        \hspace{2em}
        \includegraphics[width=\textwidth]{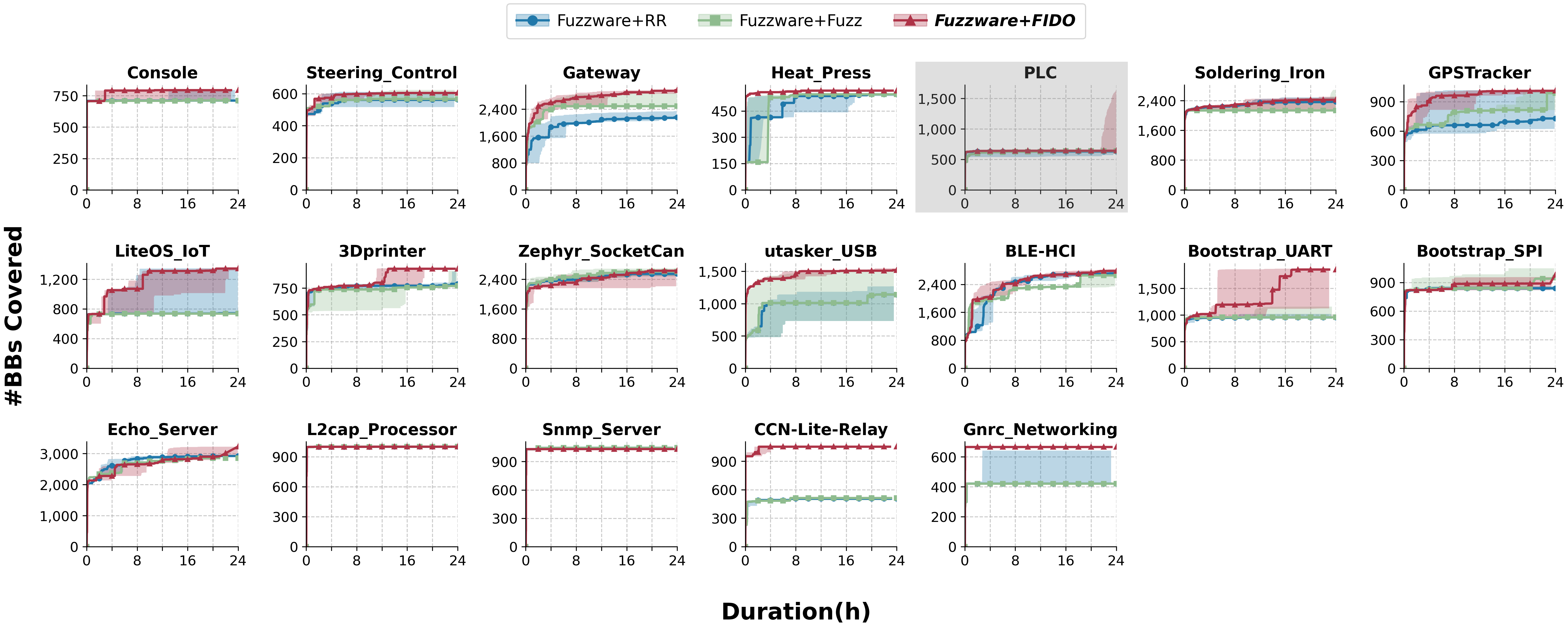} 
    \end{subfigure}
    
    \begin{subfigure}[b]{\textwidth}
        \raggedright
        \hspace{2em}
        \includegraphics[width=\textwidth]{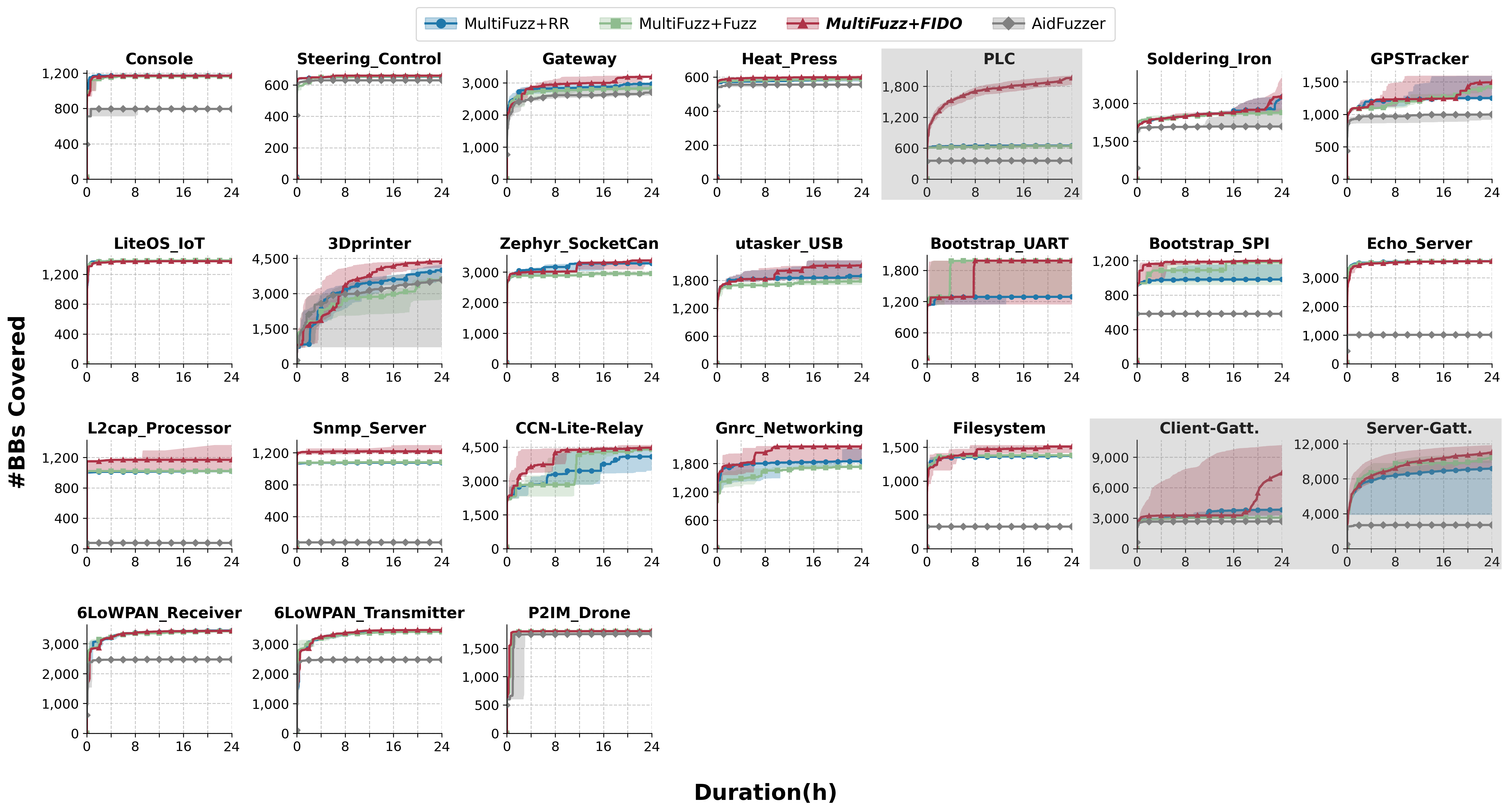} 
    \end{subfigure}
\flushleft
\footnotesize{
Note: The fuzzing process with \aid fails to start for \texttt{LiteOS\_IoT}, \texttt{Gnrc\_networking}, \texttt{$\mu$tasker\_USB}, \texttt{Zephyr\_SocketCan}, \texttt{Bootstrap(UART)}, and \texttt{BLE-HCI}.
}

\caption{Code Coverage Comparison with and without \sys on \fuzzware and \mulitfuzz, and Code Coverage of \aid}
\label{fig:cov}
\end{figure*}

%% file: tex/Meta-Review.tex
\twocolumn
\newpage 

\section{Meta-Review}

The following meta-review was prepared by the program committee for the 2026 IEEE Symposium on Security and Privacy (S\&P) as part of the review process as detailed in the call for papers.

\subsection{Summary}
This paper presents \sys (Fuzzing Input Delivery Optimizer), an extension for emulation-based firmware fuzzing frameworks that improves the delivery of fuzzing inputs in interrupt-driven embedded firmware. The approach models firmware input handling as ``CRP input routes'' consisting of availability Checks, data Retrieval, and Processing steps. FIDO uses a combination of static and dynamic analysis to identify appropriate delivery points and infer input-length bounds. Based on this model, FIDO schedules and distributes inputs across interrupt routes to avoid common issues such as data loss, starvation, or excessive input injection that limit existing fuzzers. The system is implemented as an extension for \fuzzware, \mulitfuzz, and \semu and evaluated on unit tests and 22 real-world firmware images, where it significantly improves code coverage and bug discovery, leading to the identification of several previously unknown vulnerabilities.

\subsection{Scientific Contributions}
\begin{itemize}
\item Creates a New Tool to Enable Future Science.
\item Provides a Valuable Step Forward in an Established Field.
\end{itemize}

\subsection{Reasons for Acceptance}
\begin{enumerate}
\item The paper provides a valuable step forward in an established field. It addresses the problem of input delivery in interrupt-driven firmware fuzzing, demonstrating that the timing and quantity of injected inputs significantly influence coverage and bug discovery. By introducing a novel approach orthogonal to traditional improvements in test-case generation or execution backends, \sys advances the state of the art.
\item The paper creates a new tool to enable future science. The authors implement FIDO as a practical extension compatible with existing firmware fuzzers such as \fuzzware, \mulitfuzz, and \semu, showing measurable improvements in coverage and vulnerability discovery. By planning to release \sys as open source, the work provides the research community with a reusable tool that can support future studies and be combined with other fuzzing advancements.
\end{enumerate}




%% file: bibs/main.bib
@String{Computing = "Computing" }

@String{Computer = "{IEEE} Computer" }

@String{Springer = "Springer-Verlag" }

@misc{ghidra-server,
  title={{Ghidra-Server.org provides a collaboration server on the internet for the software reverse engineering}},
  author={National Security Agency},
  howpublished={\url{https://www.ghidra-server.org/}},
  month={April},
  year={2025}
}

@inproceedings {tobias2022fuzzware,
title = {Fuzzware: Using Precise {MMIO} Modeling for Effective Firmware Fuzzing},
author={Tobias Scharnowski and Nils Bars and Moritz Schloegel and Eric Gustafson and Marius Muench and Giovanni Vigna and Christopher Kruegel and Thorsten Holz and Ali Abbasi},
booktitle = {31st USENIX Security Symposium (USENIX Security 22)},
year = {2022},
address = {Boston, MA},
url = {https://www.usenix.org/conference/usenixsecurity22/presentation/scharnowski},
publisher = {USENIX Association},

}

@inproceedings{zhou2022your,
  title={What Your Firmware Tells You Is Not How You Should Emulate It: A Specification-Guided Approach for Firmware Emulation},
  author={Zhou, Wei and Zhang, Lan and Guan, Le and Liu, Peng and Zhang, Yuqing},
  booktitle={Proceedings of the 2022 ACM SIGSAC Conference on Computer and Communications Security},
  pages={3269--3283},
  year={2022}
}

@inproceedings{won2022what,
	title = {What You See is Not What You Get: Revealing Hidden Memory Mapping for Peripheral Modeling},
	url = {https://doi.org/10.1145/3545948.3545957},
	doi = {10.1145/3545948.3545957},
	pages = {200--213},
	booktitle = {25th International Symposium on Research in Attacks, Intrusions and Defenses, {RAID} 2022, Limassol, Cyprus, October 26-28, 2022},
	publisher = {{ACM}},
	author = {Won, Jun Yeon and Wen, Haohuang and Lin, Zhiqiang},
    year = {2022},
}

@inproceedings{chong2024afriend,
  title={A Friend's Eye is A Good Mirror: Synthesizing MCU Peripheral Models from Peripheral Driver},
  author={Chongqing, Lei and Zhen, Ling and Yue, Zhang and Yan, Yang and Junzhou, Luo and Xinwen, Fu},
  booktitle={USENIX Security},
  year={2024},
}

@inproceedings{farrelly2023ember,
  title={Ember-IO: Effective Firmware Fuzzing with Model-Free Memory Mapped IO},
  author={Farrelly, Guy and Chesser, Michael and Ranasinghe, Damith C},
  booktitle = {Proceedings of the 2023 ACM Asia conference on computer and communications security},
  year={2023}
}

@inproceedings{chesser2023icicle,
  title={Icicle: A re-designed emulator for grey-box firmware fuzzing},
  author={Chesser, Michael and Nepal, Surya and Ranasinghe, Damith C},
  booktitle={Proceedings of the 32nd ACM SIGSOFT International Symposium on Software Testing and Analysis},
  pages={76--88},
  year={2023}
}

@inproceedings{chen2022sfuzz,
  title={Sfuzz: Slice-based fuzzing for real-time operating systems},
  author={Chen, Libo and Cai, Quanpu and Ma, Zhenbang and Wang, Yanhao and Hu, Hong and Shen, Minghang and Liu, Yue and Guo, Shanqing and Duan, Haixin and Jiang, Kaida and others},
  booktitle={Proceedings of the 2022 ACM SIGSAC Conference on Computer and Communications Security},
  pages={485--498},
  year={2022}
}

@misc{FreeRTOS,
   author ={Amazon Web Services},
   Title = {FreeRTOS Real-time operating system for microcontrollers},
   year ={2020},
   howpublished = {\url{https://www.freertos.org/}},
}

@misc{Arduino,
   author = {Arduino},
   Title = {Enable anyone to enhance their lives through accessible electronics and digital technologies},
   year ={2020},
   howpublished = {\url{https://www.arduino.cc/}},
  
}

@misc{RIOT,
   Title = {The friendly Operating System for the Internet of Things: RIOT},
   author = {RIOT},
   year ={2020},
   howpublished = {\url{https://www.riot-os.org/}},
  
}

@inproceedings{fengp2020p2im,
  title={P2IM: Scalable and Hardware-independent Firmware Testing via Automatic Peripheral Interface Modeling},
  author={Feng, Bo and Mera, Alejandro and Lu, Long},
  booktitle={Proceedings of Usenix Security Symposium},
  year={2020},
}

@inproceedings {scharnowski2023hoedur,
    title = {Hoedur: Embedded Firmware Fuzzing using Multi-Stream Inputs},
    booktitle = {32nd USENIX Security Symposium (USENIX Security 23)},
    year = {2023},
    address = {Boston, MA},
    url = {https://www.usenix.org/conference/usenixsecurity23/presentation/scharnowski},
    publisher = {USENIX Association},
    author={Scharnowski, Tobias and Woerner, Simon and Buchmann, Felix and Bars, Nils and Schloegel, Moritz and Holz, Thorsten},
    month = aug,
}

@inproceedings{gustafson2019toward,
  title={Toward the Analysis of Embedded Firmware through Automated Re-hosting},
  author={Gustafson, Eric and Muench, Marius and Spensky, Chad and Redini, Nilo and Machiry, Aravind and Fratantonio, Yanick and Balzarotti, Davide and Francillon, Aur{\'e}lien and Choe, Yung Ryn and Kruegel, Christophe and others},
  booktitle={22nd International Symposium on Research in Attacks, Intrusions and Defenses ($\{$RAID$\}$ 2019)},
  pages={135--150},
  year={2019}
}

@inproceedings{muench2018avatar2,
  title={Avatar2: A multi-target orchestration platform},
  author={Muench, Marius and Nisi, Dario and Francillon, Aur{\'e}lien and Balzarotti, Davide},
  booktitle={Proc. Workshop Binary Anal. Res.(Colocated NDSS Symp.)},
  volume={18},
  pages={1--11},
  year={2018}
}

@inproceedings{talebi2018charm,
  title={Charm: Facilitating dynamic analysis of device drivers of mobile systems},
  author={Talebi, Seyed Mohammadjavad Seyed and Tavakoli, Hamid and Zhang, Hang and Zhang, Zheng and Sani, Ardalan Amiri and Qian, Zhiyun},
  booktitle={27th USENIX Security Symposium},
  pages={291--307},
  year={2018}
}

@article{zhu2022fuzzing,
  title={Fuzzing: a survey for roadmap},
  author={Zhu, Xiaogang and Wen, Sheng and Camtepe, Seyit and Xiang, Yang},
  journal={ACM Computing Surveys (CSUR)},
  volume={54},
  number={11s},
  pages={1--36},
  year={2022},
  publisher={ACM New York, NY}
}

@inproceedings{liu2024co3,
  title={$\{$CO3$\}$: Concolic Co-execution for Firmware},
  author={Liu, Changming and Mera, Alejandro and Kirda, Engin and Xu, Meng and Lu, Long},
  booktitle={33rd USENIX Security Symposium (USENIX Security 24)},
  pages={5591--5608},
  year={2024}
}

@misc{Zephyr,
   Title = {{Deliver the best-in-class RTOS}},
   year ={2020},
   author = {Zephyr},
   howpublished = {\url{https://www.zephyrproject.org/}},
   note = {Last accessed: 2022-08-01}

}

@misc{SocketCan,
   Title = {{Socket CAN Sample}},
   year ={2020},
   author = {Zephyr},
   howpublished = {\url{https://docs.zephyrproject.org/latest/samples/net/sockets/can/README.html}}, 
   note = {Last accessed: 2022-05-01}

}

@misc{nxp,
   Title = {{Kinetis K64 Reference Manual}},
   year ={2014},
   author = {NXP},
   howpublished = {\url{https://www.nxp.com/webapp/Download?colCode=K64P144M120SF5RM}}, 
   note = {Last accessed: 2022-05-01}
  
}

@inproceedings{spensky2021conware,
  title={Conware: Automated Modeling of Hardware Peripherals},
  author={Spensky, Chad and Machiry, Aravind and Redini, Nilo and Unger, Colin and Foster, Graham and Blasband, Evan and Okhravi, Hamed and Kruegel, Christopher and Vigna, Giovanni},
  booktitle={Proceedings of the 2021 ACM Asia Conference on Computer and Communications Security},
  pages={95--109},
  year={2021}
}

@inproceedings{zhou2021automatic,
  title={Automatic Firmware Emulation through Invalidity-guided Knowledge Inference},
  author={Zhou, Wei and Guan, Le and Liu, Peng and Zhang, Yuqing},
  booktitle={30th USENIX Security Symposium (USENIX Security 21)},
  year={2021}
}

@inproceedings{kim2021pasan,
  title={$\{$PASAN$\}$: Detecting peripheral access concurrency bugs within $\{$Bare-Metal$\}$ embedded applications},
  author={Kim, Taegyu and Kumar, Vireshwar and Rhee, Junghwan and Chen, Jizhou and Kim, Kyungtae and Kim, Chung Hwan and Xu, Dongyan and Tian, Dave Jing},
  booktitle={30th USENIX Security Symposium (USENIX Security 21)},
  pages={249--266},
  year={2021}
}

@inproceedings{shoshitaishvili2015firmalice,
  title={Firmalice-automatic detection of authentication bypass vulnerabilities in binary firmware.},
  author={Shoshitaishvili, Yan and Wang, Ruoyu and Hauser, Christophe and Kruegel, Christopher and Vigna, Giovanni},
  booktitle={NDSS},
  volume={1},
  pages={1--1},
  year={2015}
}

@inproceedings{chen2021sharing,
  title={Sharing more and checking less: Leveraging common input keywords to detect bugs in embedded systems},
  author={Chen, Libo and Wang, Yanhao and Cai, Quanpu and Zhan, Yunfan and Hu, Hong and Linghu, Jiaqi and Hou, Qinsheng and Zhang, Chao and Duan, Haixin and Xue, Zhi},
  booktitle={30th USENIX Security Symposium (USENIX Security 21)},
  pages={303--319},
  year={2021}
}

@inproceedings{farrelly2023splits,
  title={SplITS: Split Input-to-State Mapping for Effective Firmware Fuzzing},
  author={Farrelly, Guy and Quirk, Paul and Kanhere, Salil S and Camtepe, Seyit and Ranasinghe, Damith C},
  booktitle={European Symposium on Research in Computer Security},
  pages={290--310},
  year={2023},
  organization={Springer}
}

@inproceedings{clements2020halucinator,
  title={$\{$HALucinator$\}$: Firmware re-hosting through abstraction layer emulation},
  author={Clements, Abraham A and Gustafson, Eric and Scharnowski, Tobias and Grosen, Paul and Fritz, David and Kruegel, Christopher and Vigna, Giovanni and Bagchi, Saurabh and Payer, Mathias},
  booktitle={29th USENIX Security Symposium (USENIX Security 20)},
  pages={1201--1218},
  year={2020}
}

@inproceedings{muench2018avatar,
  title={Avatar 2: A multi-target orchestration platform},
  author={Muench, Marius and Nisi, Dario and Francillon, Aur{\'e}lien and Balzarotti, Davide},
  booktitle={Proc. Workshop Binary Anal. Res.(Colocated NDSS Symp.)},
  volume={18},
  pages={1--11},
  year={2018}
}

@inproceedings{zaddach2014avatar,
  title={AVATAR: A Framework to Support Dynamic Security Analysis of Embedded Systems' Firmwares.},
  author={Zaddach, Jonas and Bruno, Luca and Francillon, Aurelien and Balzarotti, Davide and others},
  booktitle={NDSS},
  volume={14},
  pages={1--16},
  year={2014}
}

@misc{ghidra,
    author = {National Security Agency},
    Title = {{Ghidra}},
    year = {2023},
    howpublished = {\url{https://ghidra-sre.org/}},
    note = {Last accessed: 2024-11-1}
}

@article{feng2023aim,
  title={AIM: Automatic Interrupt Modeling for Dynamic Firmware Analysis},
  author={Feng, Bo and Luo, Meng and Liu, Changming and Lu, Long and Kirda, Engin},
  journal={IEEE Transactions on Dependable and Secure Computing},
  year={2023},
  publisher={IEEE}
}

@inproceedings{chesser2024multifuzz,
    author = {Chesser, Michael and Nepal, Surya and Ranasinghe, Damith C},
    title = {MULTIFUZZ: A Multi-Stream Fuzzer For Testing Monolithic Firmware},
    booktitle = {33rd USENIX Security Symposium (USENIX Security)},
    year = {2024},
}

@inproceedings{chesser2024ffxe,
    author = {Ryan, Tsang and Asmita and Doreen, Joseph and Soheil, Salehi and Prasant, Mohapatra and Houman, Homayoun},
    title = {FFXE: Dynamic Control Flow Graph Recovery for Embedded Firmware Binaries},
    booktitle = {33rd USENIX Security Symposium (USENIX Security)},
    year = {2024},
}

@article{zhu2021constructing,
  title={Constructing more complete control flow graphs utilizing directed gray-box fuzzing},
  author={Zhu, Kailong and Lu, Yuliang and Huang, Hui and Yu, Lu and Zhao, Jiazhen},
  journal={Applied Sciences},
  volume={11},
  number={3},
  pages={1351},
  year={2021},
  publisher={MDPI}
}

@inproceedings{li2022muafl,
  title={$\mu$AFL: non-intrusive feedback-driven fuzzing for microcontroller firmware},
  author={Li, Wenqiang and Shi, Jiameng and Li, Fengjun and Lin, Jingqiang and Wang, Wei and Guan, Le},
  booktitle={Proceedings of the 44th International Conference on Software Engineering},
  pages={1--12},
  year={2022}
}

@misc{PLC1descrip,
    author = {Tobias Scharnowski},
    Title = {{P2IM PLC Bug in Process\_FC1 Description}},
    year = {2022},
    howpublished = {\url{https://github.com/fuzzware-fuzzer/fuzzware-experiments/tree/main/04-crash-analysis/14}},
    note = {Last accessed: 2024-03-01}
}

@misc{PLC2descrip,
    author = {Tobias Scharnowski},
    Title = {{P2IM PLC Bug in Process\_FC3 Description}},
    year = {2022},
    howpublished = {\url{https://github.com/fuzzware-fuzzer/fuzzware-experiments/tree/main/04-crash-analysis/15}},
    note = {Last accessed: 2024-03-01}
}

@misc{PLC3descrip,
    author = {Tobias Scharnowski},
    Title = {{P2IM PLC Bug in Process\_FC15 Description}},
    year = {2022},
    howpublished = {\url{https://github.com/fuzzware-fuzzer/fuzzware-experiments/tree/main/04-crash-analysis/16}},
    note = {Last accessed: 2024-03-01}
}

@misc{PLC4descrip,
    author = {Tobias Scharnowski},
    Title = {{P2IM PLC Bug in Process\_FC16 Description}},
    year = {2022},
    howpublished = {\url{https://github.com/fuzzware-fuzzer/fuzzware-experiments/tree/main/04-crash-analysis/17}},
    note = {Last accessed: 2024-03-01}
}

@misc{GPSTracker1descrip,
    author = {Chesser, Michael and Nepal, Surya and Ranasinghe, Damith C},
    Title = {{Stdio initialization race}},
    year = {2024},
    howpublished = {\url{https://github.com/MultiFuzz/MultiFuzz-benchmarks/blob/main/crash-analysis.md#return-value-of-strtok-not-checked-for-null-in-gsm_get_imei}},
    note = {Last accessed: 2024-03-01}
}

@misc{GPSTracker2descrip,
    author = {Chesser, Michael and Nepal, Surya and Ranasinghe, Damith C},
    Title = {{Stdio initialization race}},
    year = {2024},
    howpublished = {\url{https://github.com/MultiFuzz/MultiFuzz-benchmarks/blob/main/crash-analysis.md#return-value-of-strstr-not-checked-for-null-in-sms_check}},
    note = {Last accessed: 2024-03-01}
}

@misc{GPSTracker3descrip,
    author = {Chesser, Michael and Nepal, Surya and Ranasinghe, Damith C},
    Title = {{Stdio initialization race}},
    year = {2024},
    howpublished = {\url{https://github.com/MultiFuzz/MultiFuzz-benchmarks/blob/main/crash-analysis.md#return-value-of-strstr-not-checked-for-null-in-gsm_get_time}},
    note = {Last accessed: 2024-03-01}
}

@misc{Gateway2descrip,
    author = {Chesser, Michael and Nepal, Surya and Ranasinghe, Damith C},
    Title = {{Incorrect handling of zero length sysex messages}},
    year = {2024},
    howpublished = {\url{https://github.com/MultiFuzz/MultiFuzz-benchmarks/blob/main/crash-analysis.md\#incorrect-handling-of-zero-length-sysex-messages}},
    note = {Last accessed: 2024-03-01}
}

@misc{Gateway3descrip,
    author = {Tobias Scharnowski},
    Title = {{P2IM Gateway OOB write in HAL Description}},
    year = {2022},
    howpublished = {\url{https://github.com/fuzzware-fuzzer/fuzzware-experiments/tree/main/04-crash-analysis/12}},
    note = {Last accessed: 2024-03-01}
}

@misc{Gateway4descrip,
    author = {Tobias Scharnowski},
    Title = {{P2IM Gateway OOB write in HAL Description}},
    year = {2022},
    howpublished = {\url{https://github.com/fuzzware-fuzzer/fuzzware-experiments/tree/main/04-crash-analysis/21}},
    note = {Last accessed: 2024-03-01}
}

@misc{Gateway6descrip,
    author = {Tobias Scharnowski},
    Title = {{P2IM Gateway OOB write in HAL Description}},
    year = {2022},
    howpublished = {\url{https://github.com/fuzzware-fuzzer/fuzzware-experiments/tree/main/04-crash-analysis/23}},
    note = {Last accessed: 2024-03-01}
}

@misc{HeatPressdescrip,
    author = {Tobias Scharnowski},
    Title = {{P2IM Heat\_Press Bug in get\_FC3 Description}},
    year = {2022},
    howpublished = {\url{https://github.com/fuzzware-fuzzer/fuzzware-experiments/tree/main/04-crash-analysis/13}},
    note = {Last accessed: 2024-03-01}
}

@misc{CCN-Lite-Relay1descrip,
    author = {Chesser, Michael and Nepal, Surya and Ranasinghe, Damith C},
    Title = {{Stdio initialization race}},
    year = {2024},
    howpublished = {\url{https://github.com/MultiFuzz/MultiFuzz-benchmarks/blob/main/crash-analysis.md\#stdio-initialization-race}},
    note = {Last accessed: 2024-03-01}
}

@misc{CCN-Lite-Relay2descrip,
    author = {Chesser, Michael and Nepal, Surya and Ranasinghe, Damith C},
    Title = {{Null Pointer Dereference in ccnl\_cs}},
    year = {2024},
    howpublished = {\url{https://github.com/MultiFuzz/MultiFuzz-benchmarks/blob/main/crash-analysis.md\#issue-with--encoded-characters-in-ccnl\_cs}},
    note = {Last accessed: 2024-03-01}
}

@misc{CCN-Lite-Relay3descrip,
    author = {Chesser, Michael and Nepal, Surya and Ranasinghe, Damith C},
    Title = {{Reinitialization of shared global timer}},
    year = {2024},
    howpublished = {\url{https://github.com/MultiFuzz/MultiFuzz-benchmarks/blob/main/crash-analysis.md\#reinitialization-of-shared-global-timer}},
    note = {Last accessed: 2024-03-01}
}

@misc{CCN-Lite-Relay4descrip,
    author = {Chesser, Michael and Nepal, Surya and Ranasinghe, Damith C},
    Title = {{Use After Free in evtimer struct}},
    year = {2024},
    howpublished = {\url{https://github.com/MultiFuzz/MultiFuzz-benchmarks/blob/main/crash-analysis.md\#missing-removal-from-evtimer-struct}},
    note = {Last accessed: 2024-03-01}
}

@misc{Zephyr_SocketCan2descrip,
    author = {Chesser, Michael and Nepal, Surya and Ranasinghe, Damith C},
    Title = {{Stdio initialization race}},
    year = {2024},
    howpublished = {\url{https://github.com/MultiFuzz/MultiFuzz-benchmarks/blob/main/crash-analysis.md#net-pkt-command-dereferences-a-user-provided-pointer}},
    note = {Last accessed: 2024-03-01}
}

@misc{10064descrip,
    author = {Linux Foundation},
    Title = {{CVE-2020-10064 Description}},
    year = {2020},
    howpublished = {\url{https://docs.zephyrproject.org/latest/security/vulnerabilities.html\#cve-2020-10064}},
    note = {Last accessed: 2024-03-01}
}

@misc{10065descrip,
    author = {Linux Foundation},
    Title = {{CVE-2020-10065 Description}},
    year = {2020},
    howpublished = {\url{https://docs.zephyrproject.org/latest/security/vulnerabilities.html\#cve-2020-10065}},
    note = {Last accessed: 2024-03-01}
}

@misc{10066descrip,
    author = {Linux Foundation},
    Title = {{CVE-2020-10066 Description}},
    year = {2020},
    howpublished = {\url{https://docs.zephyrproject.org/latest/security/vulnerabilities.html\#cve-2020-10066}},
    note = {Last accessed: 2024-03-01}
}

@misc{3319descrip,
    author = {Linux Foundation},
    Title = {{CVE-2021-3319 Description}},
    year = {2021},
    howpublished = {\url{https://github.com/zephyrproject-rtos/zephyr/security/advisories/GHSA-94jg-2p6q-5364}},
    note = {Last accessed: 2024-03-01}
}

@misc{3320descrip,
    author = {Linux Foundation},
    Title = {{CVE-2021-3320 Description}},
    year = {2021},
    howpublished = {\url{https://docs.zephyrproject.org/latest/security/vulnerabilities.html\#cve-2021-3320}},
    note = {Last accessed: 2024-03-01}
}

@misc{3329descrip,
    author = {Linux Foundation},
    Title = {{CVE-2021-3329 Description}},
    year = {2021},
    howpublished = {\url{https://github.com/zephyrproject-rtos/zephyr/issues/39549}},
    note = {Last accessed: 2024-03-01}
}

@misc{12140descrip,
    author = {Tobias Scharnowski},
    Title = {{CVE-2020-12140 Description}},
    year = {2022},
    howpublished = {\url{https://github.com/fuzzware-fuzzer/fuzzware-experiments/blob/main/03-fuzzing-new-targets/bug-details/CVE-2020-12140-Contiki-NG-l2cap-frame-size.md}},
    note = {Last accessed: 2024-03-01}
}

@misc{12141descrip,
    author = {Tobias Scharnowski},
    Title = {{CVE-2020-12141 Description}},
    year = {2022},
    howpublished = {\url{https://github.com/fuzzware-fuzzer/fuzzware-experiments/blob/main/03-fuzzing-new-targets/bug-details/CVE-2020-12141-Contiki-NG-SNMP-string-decode.md}},
    note = {Last accessed: 2024-03-01}
}

@misc{Mbedblerepo,
   Title = {Official Mbed BLE Examples},
   author = {Mbed OS.},
   howpublished = {\url{https://github.com/ARMmbed/mbed-os-example-ble}}, 
}

@misc{Zephyrblerepo,
   Title = {BLE HCI UART Example},
   author = {Zephyr.},
   howpublished = {\url{https://docs.zephyrproject.org/latest/samples/bluetooth/hci_uart/README.html#bluetooth_hci_uart}}, 
}

@misc{israsap,
   author = {Phil Koopman},
   Title = {Why Short Interrupt Service Routines Matter},
   note = {Last accessed: 2025-06-01},  
   year = {2025},
   howpublished = {\url{https://betterembsw.blogspot.com/2013/04/why-short-interrupt-service-routines.html}}, 
}

@misc{wikiiomode,
   author = {WikeBooks},
   Title = {Embedded Systems/IO Programming},
   note = {Last accessed: 2025-07-01},  
   year = {2025},
   howpublished = {\url{https://en.wikibooks.org/wiki/Embedded_Systems/IO_Programming}}, 
}

@inproceedings{mera2024shift,
  title={$\{$SHiFT$\}$: Semi-hosted Fuzz Testing for Embedded Applications},
  author={Mera, Alejandro and Liu, Changming and Sun, Ruimin and Kirda, Engin and Lu, Long},
  booktitle={33rd USENIX Security Symposium (USENIX Security 24)},
  pages={5323--5340},
  year={2024}
}

@inproceedings{wangaidfuzzer,
  title={AidFuzzer: Adaptive Interrupt-Driven Firmware Fuzzing via Run-Time State Recognition},
  author={Wang, Jianqiang and Wang, Qinying and Scharnowski, Tobias and Shi, Li and Woerner, Simon and Holz, Thorsten},
  booktitle={34rd USENIX Security Symposium (USENIX Security 25)},
  year={2025}
}

@inproceedings{scharnowski2025gdma,
  title={GDMA: Fully Automated DMA Rehosting via Iterative Type Overlays},
  author={Scharnowski, Tobias and Hoffmann, Simeon and Bley, Moritz and W{\"o}rner, Simon and Klischies, Daniel and Buchmann, Felix and Tippenhauer, Nils Ole and Holz, Thorsten and Muench, Marius},
  booktitle={34rd USENIX Security Symposium (USENIX Security 25)},
  year={2025}
}
